\documentclass[conference]{IEEEtran}
\pdfoutput=1
\usepackage[utf8x]{inputenc}

\usepackage[cmex10]{amsmath}
\usepackage{amsfonts}
\usepackage{algorithm}
\usepackage{algpseudocode}
\usepackage{tikz}

\usepackage[caption=false,font=footnotesize]{subfig}

\usepackage{epsfig}

\def\argmin{\mathop{\rm arg\,min}}

\newtheorem{theorem}{Theorem}[section]
\newcommand{\bTheorem}{ \begin{theorem}  }
\newcommand{\eTheorem}{ \end{theorem}    }

\newcommand{\bProof}{ \noindent {\bf Proof:} }
\newcommand{\eProof}{\hspace*{.1in} \hfill \begin{picture}(6,6)
\thicklines \put(0,0){\line(0,7){7}} \put(1,0){\line(0,7){7}}
\put(1.5,0){\line(0,7){7}} \put(2,0){\line(0,7){7}}
\put(3,0){\line(0,7){7}} \put(4.5,0){\line(0,7){7}}
\put(4,0){\line(0,7){7}} \put(5,0){\line(0,7){7}}
\end{picture} }

\def\BibTeX{{\rmfamily B\kern-.05em{\scshape i\kern-.025em b}\kern-.08em \TeX}}

\newcommand{\bEq}{ \begin{eqnarray}  }
\newcommand{\eEq}{ \end{eqnarray}    }

\newtheorem{proposition}{Proposition}[section]
\newcommand{\bProposition}{ \begin{proposition}  }
\newcommand{\eProposition}{ \end{proposition}    }

\newtheorem{Definition}{Definition}[section]
\newcommand{\bDefinition}{ \begin{Definition} }
\newcommand{\eDefinition}{ \end{Definition} }
\newcommand{\bDef}{ \begin{Definition} }
\newcommand{\eDef}{ \end{Definition} }

\newtheorem{lemma}{Lemma}[section]
\newcommand{\bLemma}{ \begin{lemma}  }
\newcommand{\eLemma}{ \end{lemma}    }

\newtheorem{Remark}{Remark}[section]
\newcommand{\bRemark}{ \begin{Remark} \rm }
\newcommand{\eRemark}{ \end{Remark}    }

\def\cN{\mathcal{N}}

\def\BibTeX{{\rmfamily B\kern-.05em{\scshape i\kern-.025em b}\kern-.08em \TeX}}

\title{Scalable Routing Easy as PIE: a Practical Isometric Embedding Protocol\\{\large Technical Report}}
\author{\IEEEauthorblockN{Julien Herzen}
 \IEEEauthorblockA{%School of Computer and Communication Sciences\\
EPFL, Lausanne, Switzerland\\
julien.herzen@epfl.ch}
\and
\IEEEauthorblockN{Cedric Westphal}
\IEEEauthorblockA{%Network Innovation Laboratory\\
Docomo Innovations, Palo Alto, CA\\
cwestphal@docomoinnovations.com}

\and
\IEEEauthorblockN{Patrick Thiran}
\IEEEauthorblockA{%Network Innovation Laboratory\\
EPFL, Lausanne, Switzerland\\
patrick.thiran@epfl.ch
}
}

\begin{document}

\maketitle

\begin{abstract}
\footnote{This work has been previously published in~\cite{herzen:pie}. The present document contains an additional optional mechanism, presented in Section~\ref{sec:asym}, to further improve performance by using route asymmetry. It also contains new simulation results.}
We present PIE, a scalable routing scheme that achieves 100\% packet delivery and low path stretch. It is easy to implement in a distributed fashion and works well when costs are associated to links. Scalability is achieved by using virtual coordinates in a space of concise dimensionality, which enables greedy routing based only on local knowledge. PIE is a general routing scheme, meaning that it works on any graph. We focus however on the Internet, where routing scalability is an urgent concern. We show analytically and by using simulation that the scheme scales extremely well on Internet-like graphs.
In addition, its geometric nature allows it to react efficiently to topological changes or failures by finding new paths in the network at no cost, yielding better delivery ratios than standard  
algorithms. The proposed routing scheme needs an amount of memory polylogarithmic in the size of the network and requires only local communication between the nodes. Although each node constructs its coordinates and routes packets locally, the path stretch remains extremely low, even lower than for centralized or less scalable state-of-the-art algorithms: PIE always finds short paths and often enough finds the shortest paths.
\end{abstract}

\section{Introduction}
% The ability to route packets from any source to any destination is probably the most fundamental functionality of the Internet. It has worked impressively well considering the scale of the networks on which it has been designed, back in the seventies. Yet, it is threatened. The scalability of routing has recently been recognized by the Internet Architecture Board as \emph{``the most important problem facing the Internet today''}~\cite{iab}. This comes from a fundamental limitation of the classic algorithms in use in the Internet (BGP, RIP, OSPF, \ldots). These algorithms find the shortest paths in the network, and they manage to do so using only local communication between the routers. They require however one entry in the routing table for each possible destination (in the Internet, for each addressable sub-network), thus requiring an amount of routing state that grows linearly with the size of the network. 
In the Internet, the tremendous growth of the number of destinations translates into a
corresponding growth of the routing tables. The Internet Architecture Board recently recognized the scalability of routing as being \emph{``the most important problem facing the Internet today''}~\cite{iab}.
The core routers need an excessive amount of resource and power to store, maintain and perform lookups in huge routing tables. The amount of traffic exchanged between the routers is proportional to the size of these tables, and the complexity of managing some state for every destination in the network
%In addition, large routing tables 
results in convergence problems and instabilities.
%In addition, they generate an always growing amount of route update messages resulting in routing instabilities: the arrival or departure of any destination potentially triggers a change at every router. 
The arrival of IPv6, along with new trends 
% such as the ``Internet of things``,
such as ubiquitous and mobile computing, 
is likely to make the number of potential destinations explode, thus exacerbating this fundamental scalability issue.
In addition, there are some other contexts where the scalability of routing can be an important concern, such as large sensor networks in which the nodes have only a very limited amount of memory.
% The scalability of routing can also be an important concern in other contexts, such as for large sensor networks, where the nodes can only afford to maintain very little state.

%Let us denote \emph{path stretch} the ratio of the length of a path found by a routing protocol, divided by the length of the shortest possible path on the graph.
There is a fundamental relationship between the size of the state required by a routing algorithm and the quality of the routes that it can find. It is well-known that to accomplish shortest path routing on any network of $n$ nodes, the routing table of each node needs to grow as $O(n)$. Indeed, if we denote by {\em path stretch} the ratio of the path length achieved by a routing protocol, divided by the shortest possible path on the graph, then it is known that any protocol that would keep the path stretch in the worst case strictly below three, would require a $O(n)$ bit state at each node as well~\cite{gavoille97}.
%The study of this relationship and the corresponding routing algorithms are the objects of \emph{compact routing}.
%An important theoretical result states that any algorithm that guarantees a path stretch strictly below three on general graphs of $n$ nodes requires routing tables of size $O(n)$~\cite{gavoille97}. 
As a direct consequence, if we want to significantly reduce the state required by routing algorithms in the future, we should consider algorithms that \emph{may} inflate the path lengths.
%Some compact routing algorithms exist that require $o(n)$ state and guarantee path stretches no bigger than three. However, a show-stopper limitation of these algorithms is that they require to process an entire view of the graph in order to devise the addresses of the nodes. This raises important practical concerns about the feasibility of such schemes for large networks. 
%Any distributed protocol implementing the algorithm would require all the nodes to re-obtain the full view of the graph after any topological change. 
% \emph{note: need to quantify precisely how ours is better in that respect}. 

%In addition, in order to provide explicit deterministic guarantees, these schemes require a state growing as $\Omega(\sqrt{n})$. While this is a great progress over $O(n)$, evidences have been given that one should aim at polylogarithmic scalability\footnote{by polylogarithmic scalability, we mean routing tables of size $O(\log^c(n))$ for some $c$.} in order to achieve \emph{``infinite scalability``} of the Internet routing system~\cite{krioukov:sigcomm}. \emph{note: other citation might be more accurate}.

%Also : compact routing mostly relies on tree routing, whereas geometric routing allows shortcuts while maintaining the delivery guarantees!

One potential avenue is to design practical protocols that create, for all the nodes of the network topology, some virtual coordinates in a metric space such that the relative position of the nodes can be expressed as a function of their distance. Greedy forwarding consists in forwarding a packet to a node's neighbor closest to the destination. As this forwarding depends only on the distances between the neighbors of a node and the destination, it is a purely local mechanism. Further, the routing table consists only of the coordinates of a node's neighbors: This information scales as the maximum degree of the graph times the size of the coordinates. These are typically of the order of $O(\log(n))$, making these so called {\em geographic} (or {\em geometric}) routing schemes very scalable ($\log(n)$ bits are already required to merely name each node in the network). In addition, as the routing decision is a simple comparison of the relative distance between a set of neighbors and a destination, the forwarding decisions are fast and easy to implement.

In his famous 1967 \emph{small-world} experiment~\cite{milgram67smallworld}, Milgram observes that human beings have the ability to efficiently route messages among themselves without having a full view of the topology; by just forwarding the messages to their acquaintances that they think are \emph{the closest to the final destination}. To some extent, the Internet and a large category of random graphs exhibit similar small-world properties~\cite{pastor-satorras04}. It is therefore natural to ask whether a more formal and explicit notion of distance can be obtained in the context of computer networks, that fits well the structure of such graphs.

%In this paper, we take advantage of the fact that 

Let $G = (V,E)$ denote the graph defined by the topology of the communication network. $V$ represents the set of nodes (routers) and $E$ denotes the set of bi-directional links connecting these nodes. Also, consider an embedding space $(X,d)$, that is the metric space $X$ equipped with the distance $d$.

For each node $v \in V$, define its set of neighbors $\cN_v$, namely: $\cN_v = \{w \in V, (v,w) \in E\}$. We recall the definition of a greedy embedding \cite{papadimitriou:conjecture}:

\bDef A greedy embedding is a mapping $f: V \rightarrow X$ such that $\forall u,w \in V, u \neq w$:
\begin{eqnarray}
\exists v \in \cN_u \mbox{ such that } d(f(v),f(w)) < d(f(u),f(w)).
\end{eqnarray}
\eDef
\vspace{0.2cm}
Applied to routing, this simply states that, if the node $u$ is trying to send or relay a packet to the destination $w$, it will always find a neighbor $v$ such that $v$ is closer to $w$ than $u$ is, and thus that delivering the packet to $v$ brings it closer to, and eventually at, its destination. Most geographical coordinate systems, including some virtual coordinate embeddings, 
do not produce greedy embeddings 
and require mechanisms to recover from local minima.

There is much theoretical work (some of which we describe in Section~\ref{sec:rw}) that considers whether a topology can be greedily embedded in a space $(X,d)$, and under which conditions. Most of this work focuses on providing guarantees, and does not lend itself to implementation, as a full view of the topology is essential to most results.
As a consequence, to our knowledge there exists no routing scheme that is practical, scalable (i.e., requiring an amount of memory polylogarithmic in $n$), achieves close to optimal path stretch and guarantees the success of routing. Our intent is to present such a scheme.

\textbf{Outline: }
% The remainder of the paper is organized as follows: 
In the next section, we summarize the related work. In Section~\ref{sec:pie}, we present PIE and the embedding protocol. In Section~\ref{sec:analysis}, we provide an analysis of PIE. 
% when applied on Internet-like graphs, 
In Section~\ref{sec:perfeval}, we present an evaluation of the performances of PIE. We discuss practical relevance for Internet routing in Section~\ref{sec:discussion} and we finally conclude in Section~\ref{sec:conclusion}.

\section{Related Work}
\label{sec:rw}
The idea of using coordinates for routing has been introduced in the context of wireless ad-hoc networks. In particular, the idea of using virtual coordinates (instead of the actual physical positions of the nodes) has been proposed as a mean to perform greedy routing without the need for a GPS receiver. \cite{rao:geographic, Dabek:vivaldi} and many others build practical schemes to create synthetic coordinates from the underlying topology. These are distributed methods, and can be implemented. However, they do not apply to all graph topologies (typically only on planar graphs) and cannot guarantee the success of greedy forwarding; the packets can be trapped in local minima.

%Face routing (see for instance, GPSR~\cite{karp:gpsr}, or the references in~\cite{Leong:techniques}) is commonly used to recover from a local minima in two dimensions. However these methods induce high congestion~\cite{subramanian:optimal}. Further, \cite{durocher:3d} demonstrates that face routing, or any deterministic, local, recovery mechanism, could not succeed in dimension higher than two. This creates the need for embeddings that are perfectly greedy in $O(\log(n))$ dimensions.
%The first propositions were to use the actual physical coordinates of the nodes on the earth~\cite{F87}. Besides the need for unpractical localization mechanisms, these methods often fail: the geographic position of the nodes might not reflect the connectivity among them, and the packets can be trapped in local minima.

Solutions such as \emph{face routing} have been proposed to guarantee the success of geographic routing when local minima are present, see for instance~\cite{goafr}.
% ~\cite{goafr, karp:gpsr}. 
% ~\cite{gfg, goafr, karp:gpsr}. 
These methods apply greedy routing by default and use a recovery mechanism when the packet is trapped in a local minimum. These deterministic recovery mechanisms only guarantee success of routing when the dimensionality of the underlying space is no more than two~\cite{durocher:3d}. In addition, backtracking out of local minima significantly inflates paths lengths and induce high congestion~\cite{subramanian:optimal}.

%In light of the drawbacks of geographic routing based on physical locations, propositions have been made to use \emph{virtual coordinates}~\cite{rao:geographic}, based only on the underlying connectivity graph. \cite{rao:geographic, Dabek:vivaldi} and others proposed practical schemes to obtain virtual coordinates in a distributed way. However, none of these schemes produces a \emph{greedy} embedding and can create local minimas.

In order to obtain greedy embeddings, it is therefore appealing to consider spaces of more than two dimensions. The fundamental tradeoff is to find a space of concise dimensionality (to guarantee scalability) that suits the embedding of a graph in a way that preserves the distances among the vertices (for routing performances). There is an ample body of theoretical work on graph embedding onto low-dimensional spaces (see~\cite{matousek} and references therein). Maymounkov~\cite{maymounkov:greedy} shows that $\log(n)$ is the minimal dimension for a Euclidean space to construct a greedy embedding of an arbitrary graph. The author also demonstrates that it is enough for trees, but his theoretical result, unfortunately, cannot be translated into a practical algorithm.

For some categories of graphs, it is possible to perform the embedding in a two dimensional Euclidean space. Indeed, Papadimitriou et al.~\cite{papadimitriou:conjecture} famously conjectured that such a space could embed any planar triangulation, and~\cite{moitra:greedy} confirms the conjecture. However, $O(n)$ bits are required to differentiate the points in the coordinate space.

Kleinberg~\cite{kleinberg:hyperbolic} and Cvetkovski et al.~\cite{Crovella:hyperbolic} consider hyperbolic spaces of 2 dimensions and~\cite{kleinberg:hyperbolic} demonstrates how to greedily embed any tree. However, here again the schemes results in coordinates of size $O(n)$ bits, and do not produce a significant gain in scalability. 
% Cvetkovski et al.~\cite{Crovella:hyperbolic} also demonstrate that graphs can be embedded in hyperbolic space, but the algorithm suffers from the same scalability issue as Kleinberg's. 
Very recently, Papadopoulos et al.~\cite{papadopoulos10} observed that uniform repartition of nodes onto a hyperbolic plane produces scale-free (Internet-like) graphs, and that the corresponding coordinates in the hyperbolic plane have desirable properties for greedy routing in these graphs. The reverse procedure has been used in~\cite{Krioukov:sustaining} to find the hyperbolic coordinates of the Internet ASs that fit the actual AS topology as well as possible. Although this work gives precious insights to understand the relations between scale-free graphs and the hyperbolic space, 
it yields an embedding that
% it 
% yields an embedding that is impractical, as it requires the whole connectivity matrix of the network in input. More importantly, it 
is not greedy and it does not provide 100\% packet delivery: routing may fail. PIE pursues similar goals but takes a different approach,
%It does not require a global view of the graph, but only local communication among the nodes. 
it does not try to fit the coordinates to a predetermined space, but lets the embedding space be determined by the topology, using only local communications between the nodes.
%However, while their finding is very interesting to understand the relations between scale-free graphs and the hyperbolic space, they consider the inverse problem. That is, they take the network topology and the node coordinates for granted and do not provide a method to find the node coordinates of an already existing network.

\cite{Pei:greedy} constructs a fully distributed practical embedding by projecting an $n$-dimensional graph topology onto a $O(\log(n))$ dimension Euclidean space using the Johnson-Lindenstrauss lemma. Despite attempting to preserve the relative
distance between points, this method is \emph{quasi}-greedy and introduces some distortion in the embedded topology, which creates local minima. It therefore requires a recovery mechanism that significantly increases the path stretch.

% There is an ample body of work on compact routing~\cite{krioukov:sigcomm,cowen:compact,krioukov:compact}.
% Thorup et al.~\cite{thorup:compact}
% show that it is possible to guarantee a maximum path stretch of three with a route table size growing as $O(\sqrt{n \log(n)})$. Indeed, while this represents a great progress over $O(n)$, a polylogarithmic growth is preferred in order to ensure \emph{''infinite scalability''} \cite{krioukov:sigcomm}. For a lower path stretch, it is known~\cite{gavoille:space,gavoille:memory} that the route table size has to be of linear size for arbitrary networks.

% Gupta et al.~\cite{gupta} and Flury et al.~\cite{greedy:flury} find a bounded stretch of 3 with $O(\log^2(n))$ coordinates for planar graphs~\cite{gupta} and combinatorial unit disk graphs~\cite{greedy:flury}.
% However, these algorithms 
% % do not work on general graphs and 
% require a full, centralized knowledge of the topology in input.

Gupta et al.~\cite{gupta} and Flury et al.~\cite{greedy:flury} find a bounded stretch
of $3$ with $O(\log^2(n))$ coordinates for planar graphs~\cite{gupta} and
combinatorial unit disk graphs~\cite{greedy:flury}. For arbitrary graphs,
the scheme of~\cite{greedy:flury} also provides a stretch of $O(\log(n))$.
However, these algorithms require a full, centralized
knowledge of the topology in input.

% Gupta et al.~\cite{gupta} find a bounded stretch of $3$ with $O(\log^2(n))$ coordinates for planar graphs.
% Flury et al.~\cite{greedy:flury} extend this result for combinatorial unit disc
% graphs, and their scheme provides a stretch of $O(\log(n))$ for arbitrary
% graphs. However, these algorithms require a full, centralized
% knowledge of the topology in input.

The idea of trading off path stretch for routing table size is the core component of the work on {\em compact routing} (see for instance~\cite{krioukov:sigcomm}). In~\cite{thorup:compact},
% ~\cite{krioukov:sigcomm,cowen:compact,krioukov:compact}. 
Thorup et al.
show that it is possible to guarantee a path stretch no larger than three with routing tables of size $O(\sqrt{n \log(n)})$. Such compact routing schemes have been successfully implemented in practice~\cite{mao:S4,singla2010}. We explore a different point in the tradeoff space, specifically, we relax the worst-case path stretch guarantee in order to provide polylogarithmic scalability, which is obviously needed to sustain any exponential growth of the Internet. We show in our evaluations that the relaxation of this guarantee does not disadvantage PIE in any way: it achieves significantly lower stretch than compact routing, and never higher than three.

%This is much higher than our polylogarithmic scheme. For a lower path stretch, it is known~\cite{gavoille:space,gavoille:memory} that the route table size has to be of linear size for arbitrary networks. 

\cite{Chen:compact} adapts the scheme of Thorup et al.\ for power-law graphs and obtains better scalability for the routing state, although still a fractional power of $n$.

\cite{brady} proposes a specialized scheme for power-law graphs, which provides polylogarithmic scalability, as PIE does. However, their method here again requires the complete topology graph in input and does not translate to a distributed protocol to build the routing tables.
In addition, it relies on \emph{tree routing}, that is, it uses only links that are spanned by some pre-constructed trees and neglect the others. PIE also constructs trees, but its geometric nature allows it to use all the links of the graph. We show in Section~\ref{sec:perfeval} that PIE finds shorter routes.
% is inherently centralized, as it requires to explicitely distinguish different regions of the graph. 

%  In this paper, Brady et al.\ present a compact routing scheme designed for power-law graphs. The routing state 
%and stretch 
% of their scheme increases polylogarithmically, as for PIE. 
% (explain better). 

% Their scheme is guaranteed to find routes whose lengths are not inflated by an \emph{additive} factor larger than $d$, while requiring $O(e \log^2n)$ state at each node, with $d$ and $e$ being parameters that depend on the network topology. In particular, $d$ and $e$ are expected to be low for power law graphs (e.g., $d \sim 10$ and $e \sim 5$ for $n = 40,000$). As the authors note, the actual values of these parameters depend on the network graph, they thus depend of statistical properties of the considered random power law graph model and therefore produce \emph{probabilistic} guarantees. They also mention that $d = O(\log n)$ for power law graphs with node degree distribution close to what is observed in the Internet. Unfortunately, while this scheme is a very interesting way of combining the theory of power law random graphs with compact routing, it is not translated into a distributed protocol.

%Table~\ref{table:sum} summarizes the fundamental properties of the state-of-the-art routing schemes.

Distributed Hash Tables (DHTs) have been used to improve the scalability of routing as well (for instance, VRR~\cite{caesar:VRR}). However, such DHTs map to source routes that require $O(\sqrt{n})$ bits to be stored on many topologies, and $O(n)$ in the worst case. \cite{ghaffari:delaunay} and references therein use Delaunay triangulations to enable greedy forwarding with bounded stretch. However, unlike our work, they assume that the nodes exist in a Euclidean space. We assume nodes in an arbitrary connectivity graph. In particular, it has been shown that Euclidean spaces are not well suited to represent Internet nodes~\cite{lee:euclidean}.

% Table~\ref{table:soa} gives a synthetic view of the contributions of PIE.
\begin{table}[htb]
    \centering
        \begin{tabular}{|l|c|c|c|c|c|}
        \hline
        & \cite{brady} & \cite{thorup:compact,mao:S4,singla2010} & \cite{Krioukov:sustaining} & \cite{Pei:greedy} & PIE\\
        \hline
	polylogarithmic scalability & \checkmark & $\times$ & \checkmark & \checkmark & \checkmark\\
	100\% success rate & \checkmark & \checkmark & $\times$ & \checkmark & \checkmark\\
	no recovery mechanism & \checkmark & \checkmark & \checkmark & $\times$ & \checkmark\\
	distributed protocol & $\times$ & \checkmark & \checkmark & \checkmark & \checkmark\\
	\hline
        \end{tabular}
	\caption{Comparison of PIE with related state-of-the-art.}
	\label{table:soa}
%         \caption{{\footnotesize Proportional utility corresponding to the throughput depicted in Figure~\ref{fig:exp_WLAN2} for the $3$-flow  scenario.}}
%                 \label{table:exp:link2} 
\end{table}
\vspace {-2em}

\section{Description of PIE}
\label{sec:pie}

\subsection{Model and Background}
\label{sec:bm}
We consider a weighted graph $G = (V,E)$ associated with a function $w : E \rightarrow \mathbb{R}_+^*$ assigning a cost to each edge of $G$. $w$ defines the usual (weighted) shortest path distance in $G$, that we denote by $d_G$. If $f : V \rightarrow X$ is an embedding of $G$ into a metric space $(X,d)$, $f$ is said to have distortion $D$ if: % need reference here???
\begin{equation*}
  \exists{}\, r > 0 \text{ such that } \forall \text{ } u,v \in V,
 \end{equation*}
 \begin{equation*}
  r \cdotp d_G(u,v) \leq d(f(u), f(v)) \leq D \cdotp r \cdotp d_G(u,v)
 \end{equation*}
An embedding with distortion 1 is said to be isometric.

We are interested in situations where the host metric space is a standard $k$-dimensional metric space $(X,d)$, where $X \in \mathbb{R}^k$, equipped with a $l_p$-norm such that $d(x,y) = \lVert x - y \rVert{}_p\text{ for all }x,y \in X \text{ and}$
\begin{equation*}
  \lVert x \rVert_p = \begin{cases}
  \sqrt[p]{\sum_{i = 1}^k \lvert x_i \rvert^p} & \text{if } 1 \leq p < \infty \text{,}
  \\
  \max_i \lvert x_i \rvert & \text{if } p = \infty \text{,}
  \end{cases}
\end{equation*}
for all $x \in X$. We denote by $l_p^k$ such a space, and thus $l_2^k$ denotes the usual $k$-dimensional Euclidean space.

%Rewrite from here, explaining the big picture with trees and scale free graphs

As there is exactly one path between any two nodes in a tree, an isometric embedding of a tree is also greedy.
Further, it is known (see Linial et al.\ \cite{linial}, Theorem 5.3) that a tree can be isometrically embedded in $l_{\infty}^{O(\log n)}$. Given these two pieces of information, we could imagine a routing scheme that first extracts a tree $T$ spanning the connection graph $G$, embeds it isometrically in $l_{\infty}^{O(\log n)}$ and uses the resulting greedy embedding of $G$ to perform greedy routing. However, this approach would not work in practice, for two main reasons: 
First, the isometric tree embedding algorithm proposed in \cite{linial} requires a full, centralized knowledge of the tree, as it recursively divides it in balanced subtrees. Second, routing over a tree is clearly inefficient because a significant number of links may not be taken into account, possibly leading to poor performance in terms of path stretch and congestion.

In the following, we address these two problems. 
% We first begin by noticing that the embedding algorithm in \cite{linial} mainly requires a global knowledge of the tree in order to guarantee that the resulting dimensionality of the embedding is of order $O(\log n)$. 
By relaxing the \emph{deterministic} guarantee on the dimensionality, we are able to devise a different, simple, isometric embedding algorithm that does not need global knowledge of the topology and is easy to implement in a distributed scenario. As shown in Section~\ref{sec:analysis}, the guarantee on the dimensionality becomes $O(\log^2n)$ \emph{with probability one} (almost surely), on the relevant categories of random graphs.

The second problem due to tree routing is addressed by constructing multiple trees with different locality levels. In such a scenario, not all the trees would span the whole graph, but most would span only a local portion of it, according to their locality level. However, the union of all the trees at each locality level would cover the whole graph. As such, each node is covered by one tree for each locality level, that is, by $\log(n)$ trees in total if we choose $\log(n)$ locality levels.

Here are the high level steps of PIE:

\begin{itemize}
 \item Extract several (rooted) trees with different locality levels from the graph, with at least one spanning the whole graph.
 \item Embed each of these trees in a separate coordinate system.
 \item When forwarding a packet, choose a tree on which to perform greedy routing and send the packet to the neighbor that provides the best progress towards the destination in this coordinate system.
%is the closest to the destination in this tree.
\end{itemize}

In the next section, we present the distributed greedy embedding algorithm in detail, using one spanning tree. The extension to several trees is explained in Section \ref{sec:multipleTrees}.

\subsection{Isometric Tree Embedding}
\label{sec:embedding}

Let $T$ denote a rooted spanning tree of $G$. We explain here how to embed $T$, and we provide later two distributed algorithms that (i) extract $T$ out of $G$ and (ii) embed $T$.
% Let us consider a rooted tree $T$ and $d_T$ the corresponding distance function in $T$. 

Let $O$ be a node of the tree $T$. At the beginning of the algorithm, $O$ is set to be the root of the tree, and the coordinate \{\texttt{0}\} is assigned to it.
%Denote $O$ the root of the tree and assign the coordinate \{\texttt{0}\} to it. 
Let us denote by $S$ the set of children of $O$ that consists of the nodes $S=\{v_0,v_1,\ldots,v_{s-1}\}$, where $s = |S|$ is the cardinality of $S$.

For each child $v_i \in S$, compute a binary representation of its index $i$. We denote by $b_i= \langle b_i^0,b_i^1,\ldots,b_i^{h-1} \rangle$ such a representation, where $h \leq \lceil \log_2 (s) \rceil$.

Let $A_i$ be the set formed by $v_i$ along with all its descendants in $T$. The algorithm appends $h$ new coordinates $\langle c_i^0,c_i^1,\ldots,c_i^{h-1} \rangle$ to all the vertices $u$ in $A_i$ as follows:
\begin{equation}
\label{eq:coordAssign2}
 c_i^j = \begin{cases}
          -d_T(u,O) \text{ if } b_i^j=0 \text{,}\\
	  d_T(u,O) \text{ if } b_i^j=1 \text{,}
         \end{cases}
\end{equation}
$0 \leq j \leq h-1$. After that, each node in $S$ plays the role of $O$, and the algorithm repeats the same procedure.
This way of assigning the coordinates goes from the root to the leaves in one pass and can be implemented in a way that induces only local communication between a node and its neighbors. In particular, at each step, the node $O$ is higher in the tree than the nodes that receive the new coordinates, and Eq.\ (\ref{eq:coordAssign2}) does not need to be evaluated for all the vertices in $A_i$ at the same time. Each node can simply infer them based on the coordinates of its parent in the tree. Therefore, each node $O$ needs only to transmit its own coordinates along with the binary map $b_i$ to each of its children $v_i$.

% The binary map $b_i$ can be any variable length binary representation of $i$ obtained with a prefix-free code. In particular, a good representation is obtained by considering all the nodes $v_i \in S$ as the leaves of a binary tree. Let $k = \lfloor \log_2(s) \rfloor$, we have $s = 2^k + r$, where $r < 2^k$.
% One can build a binary tree that has $2^k - r$ leaves at depth $k$ and $2r$ leaves at depth $k+1$. The associated binary representations of length $k$ or $k+1$ for the children $v_i \in S$ follow.
% Put differently, this corresponds to the result that can be obtained using a Huffman code to represent the $s$ children of $O$ when they are equiprobable.
The binary map $b_i$ can be any variable length binary representation of $i$ obtained with a prefix-free code. In particular, such a map of length $h \leq \lceil \log_2 (s) \rceil$ can be obtained using a Huffman code to represent the $s$ children of $O$ when they are equiprobable.

\begin{figure}
\vspace{0.1in}
\centering
\subfloat[]{%
	\includegraphics[width=.85\linewidth]{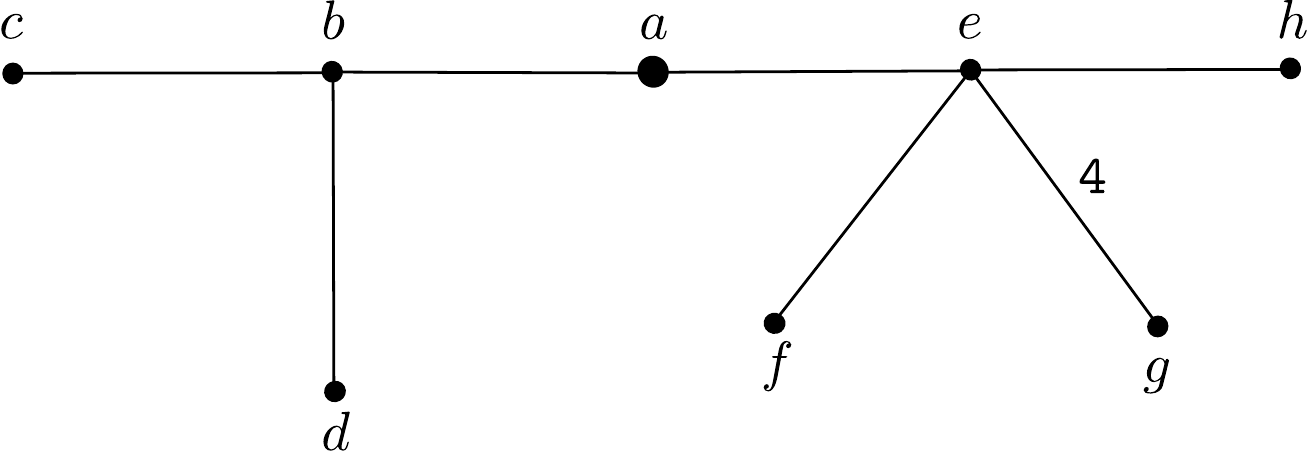}%
}\hfil
\subfloat[]{%
	\includegraphics[width=.85\linewidth]{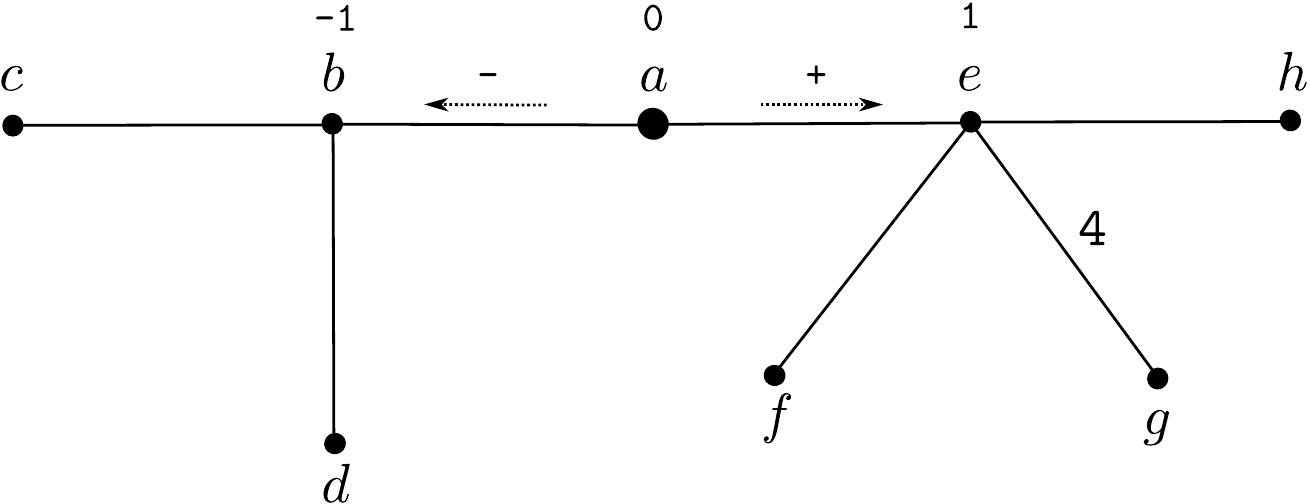}%
}\hfil
\subfloat[]{%
	\includegraphics[width=.85\linewidth]{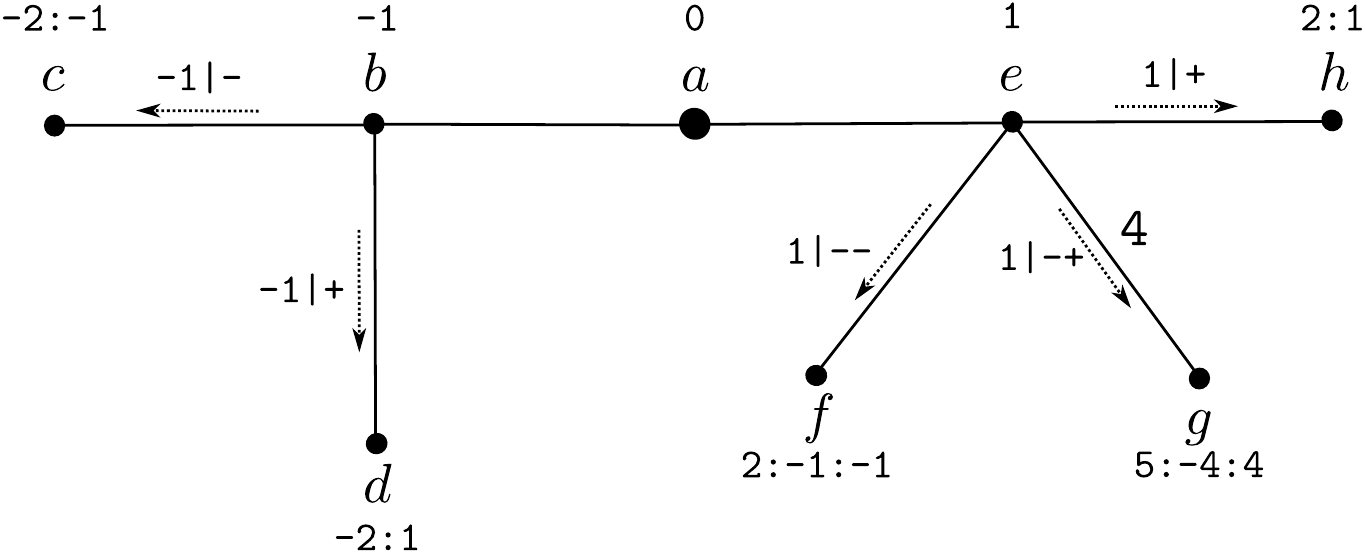}%
}
\caption{Example of isometric embedding from the root to the leaves. (a) shows the tree to embed. 
% (we show only $T$ and not the underlying graph $G$). 
The vertices are named $a$ to $h$. The root is the node $a$. We assume that all the edges have a weight of 1, except the edge ($e,g$) that has a weight of 4. (b) After having picked the coordinate \{\texttt{0}\}, the root sends the binary map (here represented by \texttt{+} and \texttt{-}) corresponding to each of its children. In (c), the nodes $b$ and $e$ play the role of $O$ and send their coordinates to each of their children, along with the corresponding binary maps. Note that the node $e$ has 3 children, resulting in a map of 2 bits for two of them and one bit for the last one. The algorithm does not require all the nodes to have the same number of coordinates, the $l_\infty$-norm is simply applied on the first common coordinates. For example, if one wants to compute the distance between the node $d$ and the node $g$, the coordinates to use are \{\texttt{-2:1}\} and \{\texttt{5:-4}\}. As $\lvert -2-5 \rvert > \lvert 1+4\rvert$, the distance is $\lvert -2-5 \rvert = 7$.}
\vspace{-1.2em}
\label{fig:example}
\end{figure}

The scheme can be slightly improved if we note that if a node $O$ is not the root and $s=1$ (i.e., it has only one child), assigning a new coordinate to all the descendants of $O$ would have no effect on their relative distance under the $l_\infty$-norm. In this case, the binary map does not need to be sent. A step-by-step example of the embedding is shown in Figure \ref{fig:example}.

The greedy forwarding procedure is straightforward: When forwarding a packet, a node considers all its neighbors that are closer to the destination and chooses the one that minimizes the overall path length (i.e., taking into account the cost of the link to go to this neighbor). Specifically, a node $v$ forwarding a packet to a destination $t$ chooses the node that satisfies:
% \vspace{-0.1in}
\begin{eqnarray}
\argmin_{u \in \cN_v \mbox{ s.t. } \lVert (t-u)  \rVert_\infty < \lVert (t-v)  \rVert_\infty} \left\{ w(v,u) +  \lVert (t-u)  \rVert_\infty \right\}.
\label{eq:forwardcost}
\nonumber
\end{eqnarray}
Note that this forwarding procedure considers \emph{all} the neighbors in $G$, and not only the neighbors in $T$. This enables shortcuts off the tree.
Indeed,
we prove in Section~\ref{sec:analysis} that this embedding of $T$ yields a greedy embedding of $G$. Therefore, this forwarding procedure always returns a next-hop closer to the destination (except if $v$ is already the destination).

% \addtolength{\textheight}{-0.1in}

\textbf{Algorithm Specification: }
The overall algorithm proceeds in two steps. First, a spanning tree is extracted from the graph and then the virtual coordinates are computed based on this tree. We specify these two steps in the form of two distinct modules, the \texttt{tree\_maintainer} and the \texttt{coordinates\_maintainer}. The \texttt{tree\_maintainer} implements a distributed spanning tree construction by using the well-known STP protocol \cite{perlman:stp}. Recall that in this protocol, each node chooses an ID and the node with the largest ID eventually becomes the root of the whole spanning tree. 
During our experiments on power-law graphs, we observed slightly better routing performances when the root was the highest degree node. We thus choose the ID to be the degree of the node (plus a small random salt to break ties if needed).
 
%Note that this procedure is completely distributed. 
%\footnote{In the practical evaluations of Section~\ref{}, this number is a function of the inverse of the degree of the node. This ensures that the node with the highest degree becomes root of the tree
%Therefore, the root is picked randomly and distributively, and it does not need to be explicitly chosen.
The STP protocol has been augmented with a straightforward improvement, in order for each node to learn who its children are when it receives messages from its neighbors. The \texttt{coordinates\_maintainer} acts separately but uses the \texttt{tree\_maintainer} in order to access the list of children.
% \vspace{0.1in}
In realistic scenarios, the topology may of course 
change over time and the links may be asynchronous. Therefore, these two modules typically act on a periodic basis in order to accommodate possible changes in the tree. The pseudo-codes corresponding to the distributed versions of the \texttt{tree\_maintainer} and \texttt{coordinates\_maintainer} are shown in Algorithms \ref{alg:simpleTreeMaint} and \ref{alg:coordMaint}, respectively. 
% The pseudo-codes are written in an object-oriented style, we use the following conventions:
We use the following conventions for the pseudo-code:
\begin{itemize}
% \vspace{-0.7em}
\item $a.b$ denotes the field $b$ of the element $a$.
\item $A[i]$ denotes the element at index $i$ of the data structure $A$ (the indexes start at $0$).
\item $A$.length denotes the number of elements in the data structure $A$.
\item treeMsg and coordMsg are the two message types used by the \texttt{tree\_maintainer} and the \texttt{coordinates\_maintainer}, respectively. When they are constructed, they receive as arguments the values of the fields that they will carry.
% \vspace{-0.7em}
\end{itemize}
The other notations should be clear from the context.
% \vspace{0.1in}
% \addtolength{\textheight}{0.1in}

% \addtolength{\topmargin}{0.08in}
\begin{algorithm}[t]
\begin{algorithmic}
\begin{small}
% \algblock[per]{Periodically}{end}

\State \textbf{Init}:

	\State $children \leftarrow \emptyset$
	\State $u.rootId \leftarrow \text{degree of }u$ (+ random salt in $[0,1[$ to break ties)
	\State $u.height \leftarrow 0$
	\State $u.parent \leftarrow \emptyset$

%\State
\vspace{0.06in}

\State \textbf{Periodically}:
	\State send treeMsg($u.rootId, u.height, u.parent$) to each neighbor of $u$

%\State
\vspace{0.06in}

\State \textbf{Upon reception of} treeMsg $msg$ from neighbor $v$:
	\State $w_v \leftarrow w(u,v)$ \Comment{cost from $u$ to $v$}
	\If{$(msg.rootId > u.rootId)$ or $(msg.rootId = u.rootId$ and $msg.height + w_v < u.height)$}
		\State $u.parent \leftarrow v$
		\State $u.height \leftarrow msg.height + w_v$
		\State $u.rootId \leftarrow msg.rootId$
	\EndIf
	\If{$msg.parent = u$ and $v \notin children$}
		\State $children$.add($v$)

	\EndIf
	\If{$msg.parent \neq u$ and $v \in children$}
		\State $children$.remove($v$)
	\EndIf

%\State
\vspace{0.06in}

\Procedure {GetChildren}{}
	\State return $children$
\EndProcedure
\end{small}
\end{algorithmic}
\caption{\texttt{tree\_maintainer} at node $u$}
\label{alg:simpleTreeMaint}
\end{algorithm}

% \vspace{-1em}

\begin{algorithm}
\caption{\texttt{coordinates\_maintainer} at node $u$}
\begin{algorithmic}
\begin{small}
% \algblock{Periodically}
\State \textbf{Init}:
	\State $u.coords[0] \leftarrow 0$ % check in details if we can simplify that...
\vspace{0.06in}

\State \textbf{Periodically}:
	\State $children \leftarrow$ \texttt{tree\_maintainer}.\Call{getChildren}{}
	\If {change in childhood}
		\State \Call{NotifyChildren}{}
	\EndIf

%\State
\vspace{0.06in}

\Procedure {NotifyChildren}{}
	\State $s \leftarrow$ number of children
	\If {$s = 1$}
		\State send coordMsg($u.coords$) to $children[0]$
	\EndIf
	\If {$s > 1$}
		\For{$i=0$ to $s-1$}
			\State $b_i \leftarrow$ prefix-free binary representation of $i$
% 			\If{$s-i \leq 2^{\lceil \log_2 s \rceil} - s$}
% 				\State $b[\lceil \log_2 s \rceil - 1] \leftarrow \emptyset$
% 			\EndIf
			\State send coordMsg($u.coords, b_i$) to $children[i]$
		\EndFor
	\EndIf
\EndProcedure

%\State
\vspace{0.06in}

\State \textbf{Upon reception of} coordMsg $msg$ from parent $p$:
	\State $w_p \leftarrow w(u,p)$ \Comment{cost from $u$ to $p$}
	\State $l \leftarrow msg.coords$.length
	\For{$k=0$ to $l-1$}
		\If {$msg.coords[k] < 0$}
			\State $u.coords[k] \leftarrow msg.coords[k] - w_p$
		\Else
			\State $u.coords[k] \leftarrow msg.coords[k] + w_p$
		\EndIf
	\EndFor

	\For{$k=0$ to $msg.b$.length$ - 1$}
		\If {$msg.b[k] = 0$}
			\State $u.coords[l+k] \leftarrow -1 \cdot w_p$
		\EndIf
		\If {$msg.b[k] = 1$}
			\State $u.coords[l+k] \leftarrow w_p$
		\EndIf
	\EndFor
	\If{$u.coords$ changed}
		\State \Call{NotifyChildren}{}
	\EndIf
\end{small}
\end{algorithmic}
\label{alg:coordMaint}

\end{algorithm}

\subsection{Extension to Several Trees}
\label{sec:multipleTrees}
This embedding is a significant improvement over tree routing. It can still be improved by using multiple trees. Building only one tree spanning the graph takes into account exactly $(n-1)$ links when computing the coordinates and it ignores all the other links.  This can lead to some sub-optimal routing decisions when the shortest path between two nodes contains a link that is not included in the spanning tree.

We now describe how to use several trees so that any given link has a high probability of being spanned by one of the trees. 
% We will see in Section~\ref{sec:perfeval} that $O(\log(n))$ trees lead to substantial performance gain and make the performances of the scheme to scale with the size of the network.
An obvious solution would be to construct multiple spanning trees. However, in order to keep a small overall number of coordinates, each node has to belong to a small number of trees. If all these trees span the whole graph and have randomly chosen roots, it is likely that some of these roots will be close to each other; this would lead to similar, redundant trees, with little or no performance gain.

% Instead, we propose a method that builds on the observation that Internet-like graphs are self-similar. It consists in partionning the graph $m$ times:
Instead, we propose to partition the graph $m$ times: 
The first partition divides the graph in two pieces, the second partition divides the graph in four pieces and, more generally, the $l$-th partition divides the graph in $2^l$ pieces. Each of these $m$ partitions defines what we denote a \emph{locality level}.
% We denote the corresponding values of $0\leq l \leq m$, the \emph{locality levels} of the graph. 
For the locality level $l$, each of the corresponding $2^l$ pieces of the graph is spanned by one tree and we denote these $2^l$ trees the trees \emph{of level} $l$.
% each of these $2^l$ trees is said to be \emph{of level} $l$. 
Note that there is only one tree of level $0$ and that it spans the whole graph.
Of course, computing such exact partitions for each of the $m$ locality levels would require a global knowledge of the graph.
% Instead, we propose to partition the graph iteratively in \emph{locality levels}. In this case, each node belongs to $m$ trees, with $m \in O(\log n)$. Each of these trees corresponds to a different locality level.
% Only the first tree (of level $0$) spans the whole graph. Those remaining span a portion of the graph whose size decreases as their level increases. Typically, we would like the graph to be partitioned in $2^l$ trees of level $l$. Of course, such an exact partition would require a global knowledge of the graph. 

In order to keep a distributed solution, we slightly modify the procedure and adopt the following election process: For each locality level $0 \leq l \leq m-1$, each node elects itself as the root of a tree of level $l$ with a probability of $2^l/n$, independently of the other nodes. We have therefore an \emph{expected} number of $2^l$ trees of level $l$, for all $0\leq l \leq m-1$. Each of the trees is then constructed in a similar way as described in the previous section; for every locality level $l$, each node chooses to belong to the tree of level $l$ whose root is the closest (breaking ties arbitrarily). Each node maintains therefore $m$ independent sets of coordinates, corresponding to the $m$ trees (of levels $0$ to $(m-1)$) to which it belongs. We will see in Section~\ref{sec:perfeval} that taking $m \in O(\log(n))$ leads to substantial performance gain and makes the performances of the scheme to scale with the size of the network.
% Instead, we propose that each node becomes the root of a tree of level $l$ with probability $2^l/n$. Each tree is constructed in a similar way as described in the previous section, except that now each node chooses to belong to the tree of level $l$ whose root is the closest (breaking the ties arbitrarily) with respect to the distance induced by the link weights $w$.

\begin{figure}[t]
\vspace{0.1in}
\centering
% \subfloat[]{%
% 	\includegraphics[width=.75\linewidth]{figures/ex2_1.pdf}%
% }\hfil
\subfloat[]{%
	\includegraphics[width=.75\linewidth]{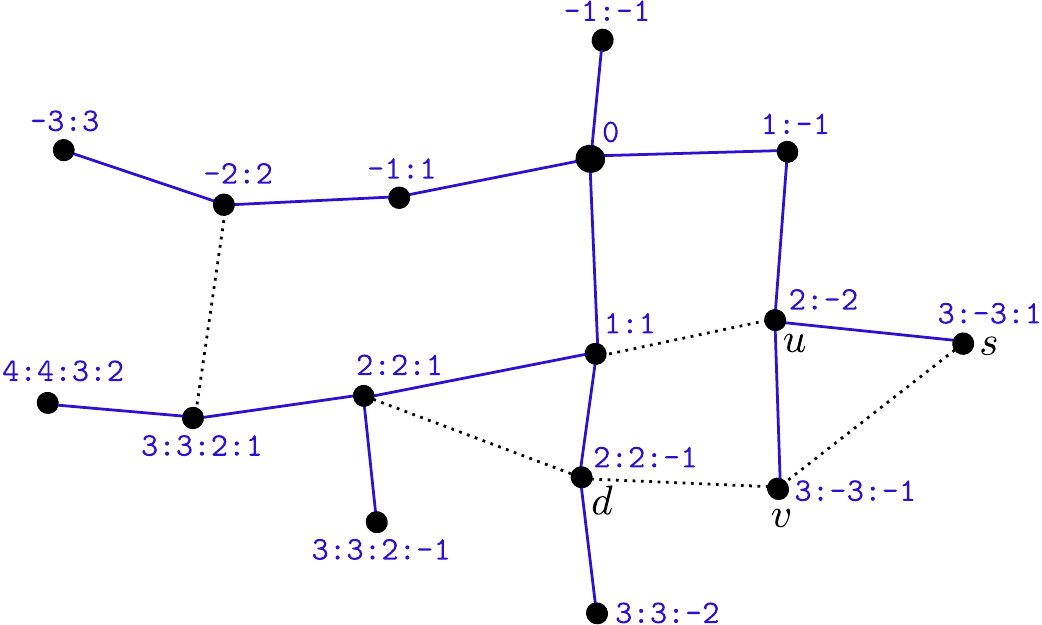}%
}\hfil
\subfloat[]{%
	\includegraphics[width=.75\linewidth]{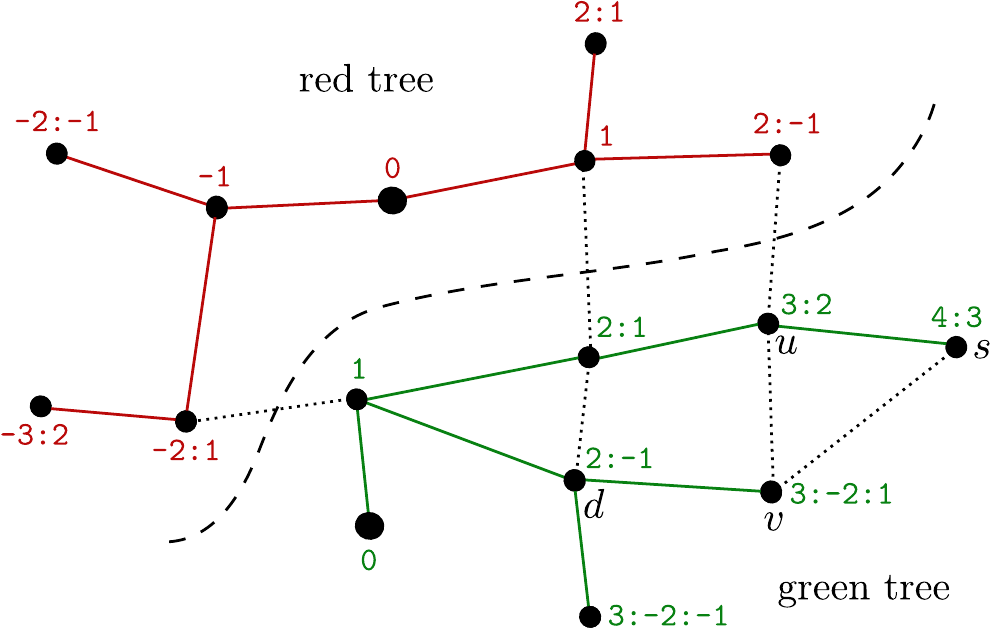}%
	\label{fig:example2b}
}
\caption{Example of an embedding using several trees. We assume for  clarity that all the edges have a cost equal to one. (a) shows a tree of level $0$ that spans the whole graph (with solid edges) and a corresponding set of coordinates at each node. (b) shows two trees of level $1$, each with solid edges, along with the corresponding coordinates, which we denote by "red" and "green". We are interested in the case where a source $s$ wants to send a message to a destination $d$. $s$ will compare the coordinates in the trees that its neighbors have in common with $d$. Here, the neighbors of $s$ are $u$ and $v$ and they both have the level 0 and the "green" level 1 sets of coordinates in common with $d$. $s$ will find that $u$ is at distance 4 of $d$ using the level 0 coordinates, and at distance 3 using the "green" coordinates. Similarly, $v$ is at distance 5 with the level 0 coordinates and 1 with the "green" ones. Therefore, $s$ will forward the packet to $v$, which is the optimal choice here. Note that if only the tree of level 0 was present, $s$ would have forwarded the packet to $u$. The "green" tree provides a valuable shortcut in this situation.}
\vspace{-1.5em}
\label{fig:example2}
\end{figure}

The only necessary condition to route all the packets successfully is that all the nodes have at least one of their $m$ sets of coordinates in common (i.e., that they belong to at least one common tree). This condition 
can be trivially satisfied by ensuring that at least one node deterministically becomes the root of a tree of level $0$. 
This node can be for instance the one having the largest ID, as in the single-tree case.
% is trivially satisfied by deterministically having at least one tree of level $0$. 
Now, when evaluating the distance between two nodes, one just chooses the $l_\infty$-norm that is minimized over the trees that the two nodes have in common.
We denote by $u_l$ the coordinates of node $u$ in the tree of level $l$ to which it belongs. In addition, we denote by $T_{u,t}^l$ a tree of level $l$ to which both the node $u$ and the node $t$ belong.
% Additionally, we write $u_i$ the coordinates of $u$ in the tree of level $i$ to which the node $u$ belongs. 
When a node $v$ wants to transmit a packet to a destination $t$, it chooses the node that satisfies:
\begin{eqnarray}
\label{eq:forwardcost2}
\argmin_{u \in \cN_v} \min_{\substack{l \mbox{ } : \mbox{ } \exists T_{u,t}^l \mbox{ s.t. }\\
 \lVert(t_l-u_l)\rVert_\infty < \lVert(t_l-v_l)\rVert_\infty}}
\left\{ w(v,u) +  \lVert (t_l-u_l)  \rVert_\infty \right\}.
\end{eqnarray}
This forwarding procedure needs to be able to uniquely identify the trees. This can easily be done using the identifier of the root, of size $\log(n)$.
% This approach still leads to a greedy embedding.
Figure \ref{fig:example2} shows an example of our embedding using several trees.

The intuition now is that a large number of small trees having a high locality level provides fine-grained coordinates for local paths, while larger trees of lower levels provide coarse-grained coordinates for the longer routes and tie everything together, much like in a divide-and-conquer strategy.
%  high number of small trees of higher levels provide fine-grained coordinates that are good 
% are likely to produce good coordinates for local paths, and larger trees of lower levels provide coordinates for the longer routes and tie everything together, much like in a divide-and-conquer strategy.

\textbf{Algorithm Specification: }
The distributed implementation simply consists in a generalization of the \texttt{tree\_maintainer} and the \texttt{coordinates\_maintainer} to use $m$ independent trees, as described above.
% To ensure that at least one tree spans the whole graph, one node (for example the one having the largest ID) becomes the root of a tree of level $0$, exactly as in the single-tree case.
% Then, 
% for $0\leq l \leq m-1$,
% each node independently chooses to become root of a tree of level $l$ with probability $2^l/n$, as described above. For each level, each node simply chooses to belong to the tree that has the closest root, breaking ties arbitrarily.

% \subsubsection{Improvement for Internet-like graphs}
% Slightly better performances can be obtained if the tree of level 0 is rooted at the highest degree node. 

\subsection{Source-aided Geometric Routing}
\label{sec:asym}
The routes found by PIE are not necessarily symmetric: When a source $s$ sends a packet to a destination $d$, the set of edges chosen by the forwarding procedure may be different than if the packet were sent by $d$ to $s$. To see this, consider again the example of Figure~\ref{fig:example2b}. The path from $s$ to $d$ goes through $v$, whereas the path from $d$ to $s$ goes through $u$ and is one hop longer.
%Therefore, the length of the paths found by PIE in the two directions $s \rightarrow d$ and $s \leftarrow d$ may be different. This suggests the following \emph{optional} extension of PIE.
Let $P_{s \rightarrow d}$ denote the path found by PIE when routing from $s$ to $d$.
It consists in the sequence of nodes through which any packet from $s$ to $d$ is routed using the geometric coordinates, including $s$ and $d$ themselves.
Similarly, let $P_{d \rightarrow s}$ denote the path from $d$ to $s$.
For a path $P$, its length can be written as $d(P) = \sum_{i=1}^{|P|-1}{w(P_{i},P_{i+1})}$, where $P_i$ is the $i$-th element of the path $P$ and $w$ is the link weight function. Note that the fact that the routes found by PIE are not necessarily symmetric implies that $d(P_{s \rightarrow d})$ is not necessarily equal to $d(P_{d \rightarrow s})$. This motivates the following \emph{optional source-routing extension} of PIE in order to use the shortest of both routes.

\begin{figure}
\centering
\includegraphics[width=.60\linewidth]{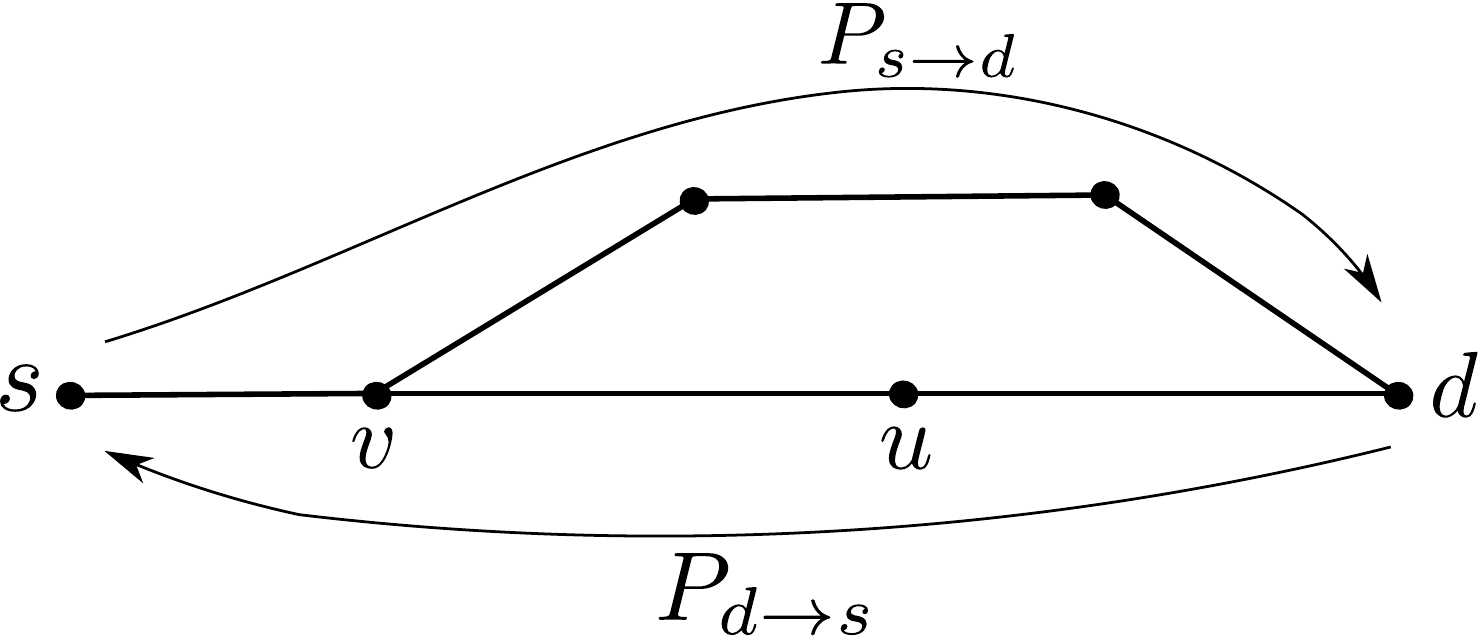}%
%\vspace{-1em}
\caption{Illustration of route asymmetry.}
%\vspace{-1.2em}
\label{fig:ex-asym}
\end{figure}

Assume that two nodes $s$ and $d$ are engaged in a bi-directional communication, which involves packets sent by $s$ to $d$ as well as packets sent by $d$ to $s$, as would for instance happen with a TCP connection.
Assume also, without loss of generality, that $s$ sends the first packet.
Our approach relies on identifying the bifurcations between $P_{s \rightarrow d}$ and $P_{d \rightarrow s}$ when the first packet in each direction is sent.
Consider an intermediate node $v$ of the return path $P_{d \rightarrow s}$, and let $u$ be the node that precedes $v$ in $P_{d \rightarrow s}$ (see Figure~\ref{fig:ex-asym}). The first packet sent by $d$ to $s$ goes through $u$ and then through $v$. When $v$ receives this packet, it checks whether $u$ would be the next hop for the destination $d$ using the virtual coordinates. If this is not the case, there is a bifurcation and the ID of $u$ is added to the packet's header. The first two packets in each direction are also used to measure $d(P_{d \rightarrow s})$ and $d(P_{s \rightarrow d})$.

When $s$ receives the first packet back from $d$, it compares the length of the two paths. If $d(P_{d \rightarrow s}) < d(P_{s \rightarrow d})$, the reverse path is shorter and $s$ stores the bifurcation(s) -- namely, a sequence of node IDs -- from the header of the packet that it just received from $d$. These bifurcations are then added to each packet that $s$ sends to $d$. Now, when receiving a packet from $s$ to $d$, each intermediate node in $P_{s \rightarrow d}$ checks if one of its neighbors belongs to the bifurcation set indicated in the packet. If a neighbor belongs to the set, the packet is forwarded to this node. Otherwise, the virtual coordinates are used. The complete procedure at the intermediate nodes is described in Algorithm~\ref{alg:fwdProcedure} and the procedure at the source $s$ is described in Algorithm~\ref{alg:sendProcedure}.

In the pseudo-code, a packet $pkt$ may contain a bifurcation set $pkt.bifSet$. This set is built during the first bi-directional exchange, and it may be used by the source $s$ if the reverse path is shorter than the direct one. The variables $pkt.dist1$ and $pkt.dist2$ are used to measure $d(P_{s \rightarrow d})$ and $d(P_{d \rightarrow s})$, respectively. 

Note that identifying the first packet in each direction only requires only one additional bit in the header. Storing the set of bifurcations may require more memory. However, we observe in Section~\ref{sec:perfeval} that the number of node IDs that compose the bifurcation set is below $1.5$ on average, at most $6$, and essentially constant with the network size.

%This procedure requires $s$ to know both $d(P_{d \rightarrow s})$ and $d(P_{s \rightarrow d})$. This is easy to achieve after one bi-directional exchange. Every intermediate node in $P_{s \rightarrow d}$ maintains  the distance traveled by the first packet and writes it in the header. When receiving this packet, $d$ knows $d(P_{s \rightarrow d})$ and reports it in the first packet it sends to $s$. The same procedure is applied by the intermediate nodes in $P_{d \rightarrow s}$, and $s$ knows both distances after receiving the first packet from $d$. Finally, each intermediate nodes needs also to know when they receive the first packet in each direction. This can simply be achieved by adding a one-bit flag to the packets' headers.

\begin{algorithm}[t]
\caption{Forwarding procedure at node $v \in P_{s \rightarrow d}$}
\begin{algorithmic}
\begin{small}
	\State \textbf{Upon reception of first packet $pkt$ with source $s$ and destination $d$:}
	\State Let $u$ be the predecessor of $v$ in $P_{s \rightarrow d}$
%	%\State Update the travelled distance:
	\State $pkt.dist1 \leftarrow pkt.dist1 + w(u,v)$
	\If {$u$ does not satisfy Expression (\ref{eq:forwardcost2}) for destination $s$}
		\State $pkt.bifSet \leftarrow pkt.bifSet \cup u$
	\EndIf
	\If {$v = d$}
		\State Store $d(P_{s \rightarrow d}) \leftarrow pkt.dist1$
		\If {$pkt.dist2 \neq $ \texttt{null}}
			\State Store $d(P_{d \rightarrow s}) \leftarrow pkt.dist2$
		\EndIf
		\State Store $bifSet_{s \rightarrow d} \leftarrow pkt.bifSet$
	\EndIf
	
	\vspace{0.06in}
	
	\State \textbf{When forwarding a subsequent packet $pkt$ from $s$ to $d$:}
	\If {$pkt.bifSet \cap v.neighbors \neq \emptyset$}
		\State Forward to the (first) node in $pkt.bifSet \cap v.neighbors$
	\Else
		\State Forward using Expression (\ref{eq:forwardcost2})
	\EndIf
\end{small}
\end{algorithmic}
\label{alg:fwdProcedure}
\end{algorithm}

\begin{algorithm}[t]
\caption{Packet emission procedure at source $s$ for destination $d$}
\begin{algorithmic}
\begin{small}
	\State \textbf{When sending the first packet $pkt$ to $d$:}
	\State set $pkt.dist1 \leftarrow 0$
	\If {$d(P_{d \rightarrow s})$ is known}
		\State set $pkt.dist2 \leftarrow d(P_{d \rightarrow s})$
	\Else
		\State set $pkt.dist2 \leftarrow\;$\texttt{null}
	\EndIf
	
	\vspace{0.06in}
	
	\State \textbf{When sending a subsequent packet $pkt$:}
	\If {$d(P_{d \rightarrow s}) < d(P_{s \rightarrow d})$}
		\State set $pkt.bifSet \leftarrow bifSet_{d \rightarrow s}$
	\EndIf
	
\end{small}
\end{algorithmic}
\label{alg:sendProcedure}
\end{algorithm}

\section{Analysis}
\label{sec:analysis}

In this section, we derive some simple facts about the scalability and performance of PIE when applied on Internet-like graphs. For simplicity, we consider only the embedding of a single spanning tree $T$ on an unweighted graph (that is, using the hop count metric). Extensions of the results to multiple trees and weighted graphs are immediate in most cases. 

\subsection{Internet-like graphs}
%Features common to most Internet models

Let us write $\text{diam}(G)$ for the diameter of a graph $G = (V,E)$. For a node $u \in V$, we write $\text{dim}(u)$ for the number of coordinates assigned to $u$ by the single tree embedding procedure of PIE. Let us also denote by $\text{dim}(G)$ the highest such number, i.e., $\text{dim}(G) = \max_{u \in V}{\text{dim}(u)}$.

It has been pointed out by several research groups
(see ~\cite{faloutsos99, mahadevan06})
% ~\cite{faloutsos99, alderson05, mahadevan06} 
that the connectivity graph of the Internet exhibits a power law node-degree distribution, both at the router and at the AS levels. In these graphs, the proportion of nodes having degree $k$ is proportional to $k^{-\lambda}$ for some constant $\lambda > 1$. For the Internet, $\lambda$ has consistently been estimated in the range $2 < \lambda < 2.3$~\cite{pastor-satorras04}. Such a degree distribution (in particular when $2 < \lambda < 3$) leads to very particular structural properties, among which the fact that the graphs typically exhibit extremely small distances between the vertices 
% ($\overline{d_G} \sim \log \log n$ and $\text{diam}(G) \sim \log n$~\cite{Chung02theaverage}),
(with $\text{diam}(G) \sim \log n$~\cite{Chung02theaverage}),  
hence the \emph{small-world} denomination. In the following, we denote by $G(n, \lambda)$ the realization of an $n$-nodes random graph such that the expected degree sequence $(k_1, k_2, ...,k_n)$ follows a power law with exponent $\lambda$ and an edge between two nodes $u_i$ and $u_j$ is created independently with probability proportional to $k_ik_j$~\cite{Chung02theaverage}.
Some authors use the term \emph{scale-free} for such graphs. As this term is not defined unambiguously and may imply some other properties that we do not need here~\cite{li:scalefree}, we only use the term \emph{power law graph}.

\subsection{Success Ratio}

\bTheorem
For any connected graph $G$, the embedding of $G$ produced by PIE ensures the success of routing.
\eTheorem
\bProof
We need to show that PIE produces a greedy embedding.
As $T$ is a subgraph of $G$ that contains all the vertices of $G$, it is clear that a greedy embedding of $T$ is also a greedy embedding of $G$. In addition, an isometric embedding of $T$ is a greedy embedding of $T$. It suffices therefore to show that PIE produces an isometric embedding of $T$. 
% For any two nodes $u,v \in T$, we have exactly one the following:
% \begin{itemize}
%  \item $d_G(u,r) = d_G(v,r)$,
%  \item $d_G(u,r) > d_G(v,r)$,
%  \item $d_G(u,r) < d_G(v,r)$.
% \end{itemize}
For any node $u \in T$, write $\langle u^0, u^1, \ldots, u^{\text{dim}(u)-1} \rangle$ its coordinates. For any two nodes $u,v \in T$, write $O_{u,v}$ their least common ancestor in $T$. 
Every node above $O_{u,v}$ in $T$ assigns the same coordinates to $u$ and $v$. $O_{u,v}$ assigns coordinates with magnitude $|d_T(O_{u,v},u)|$ to $u$ and $|d_T(O_{u,v},v)|$ to $v$ in Eq.~(\ref{eq:coordAssign2}), with at least two of these coordinates, say $u^h$ and $v^h$, having opposite signs. Therefore, $\exists h$ s.t. $|u^h - v^h| = d_T(O_{u,v},u) + d_T(O_{u,v},v) = d_T(u,v)$, because $O_{u,v}$ is the least common ancestor of $u$ and $v$. Moreover, as every node below $O_{u,v}$ in $T$ assigns coordinates with a magnitude strictly smaller than $|d_T(O_{u,v},u)|$ to $u$ and $|d_T(O_{u,v},v)|$ to $v$, $\lVert{}u - v\rVert{}_\infty = |u^h - v^h| = d_T(u,v)$.
\eProof
%Treat multiple tree case
% Therefore, $\exists$ $h$ s.t. $|u^h - v^h| = d_T(u,v)$ and $|u^i - v^i| < |u^h - v^h|$ $\forall i \neq h$. In particular, $h$ is the index of a coordinate attributed to $u$ and $v$ by $O_{u,v}$ in Equation~\ref{eq:coordAssign2}. 
% \eProof
\subsection{Scalability}
\label{sec:analysis:scal}
We give a probabilistic upper bound on the number of coordinates that are required to describe the position of the nodes in large graphs. 
% For a node $u \in G$, let us write $\text{dim}(u)$ the number of coordinates assigned to $u$ by the single tree embedding procedure of PIE. Let us also denote $\text{dim}(G)$ the highest such number, i.e., $\text{dim}(G) = \max_{u \in G}{\text{dim}(u)}$.

\bTheorem
\label{th:scal}
Let $G(n, \lambda)$ be an $n$-nodes realization of a power law graph with $2 < \lambda < 3$. We have:
\begin{equation}
\label{eq:scal}
 \text{dim}(G(n, \lambda)) \in O(\log^2(n))
\end{equation}
almost surely.
\eTheorem
\bProof
Let $r$ denote the root of $T$. For any node $u$, let $P$ be the set of all the nodes above $u$ in the unique path from $r$ to $u$ in $T$. For each node $v \in P$, the embedding algorithm assigns $\lceil \log_2(\delta_v) \rceil$ new coordinates to $u$, where $\delta_v$ denotes the degree of the node $v$ (see Eq.~(\ref{eq:coordAssign2})). Obviously, we have that $\forall v \in G, \delta_v \leq \Delta$, where $\Delta$ is the maximum node degree in $G$. In addition, as $T$ is the union of the shortest paths from $r$ to all the other nodes in $G$, we have $\lvert P \rvert = d_G(r,u)$. We can therefore write the upper bound $\text{dim}(u) \leq \lceil \log_2(\Delta)\rceil d_G(u,r)$. We have:
\begin{itemize}
 \item $\Delta < n$,
 \item $d_G(u,r) \leq \text{diam}(G) \in O(\log n)$ a.s. (\cite{Chung02theaverage} Theorem 4).
\end{itemize}
Relation (\ref{eq:scal}) follows.
\eProof

%Note that the bound is quite loose => less than log^2 in practice!

% This means that, with probability one, we embed $T$ (and thus $G$) in $l^{O(\log^2(n))}_\infty$ when $n$ is large.
This means that, with probability one, PIE embeds $T$ (and thus $G$) in $l^{O(\log^2(n))}_\infty$.
If we consider the multi-tree case, each node belongs to $O(\log n)$ trees and thus PIE almost surely embeds $G$ in $l^{O(\log^3(n))}_\infty$.
%This is to contrast with the tree embedding in $l^{O(\log(n))}_\infty$ of~\cite{linial}. We retain the polylogarithmic scalability while using an embedding that lends itself to a distributed implementation. 
Note that this bound holds for any graph with diameter $O(\log n)$. In particular, it holds for more classic random graphs (see for example~\cite{lu:diameter}).

\subsection{Performance}
%\subsubsection{Path stretch}
As a node ``knows'' all of its neighbors, the algorithm finds all the 1-hop routes with stretch $1$. Therefore, a route between a source $u$ and a destination $v$ may exhibit a stretch larger than $1$ only if $d_G(u,v) \geq 2$.

As the embedding is greedy, the longest possible route that the routing procedure can find has length $\text{diam}(T) \leq 2 \cdot \text{diam}(G)$. 
% (explain that). 
Therefore, the worst possible stretch is $\text{diam}(G)$. Using Theorem 4 in~\cite{Chung02theaverage}, we have just shown the following:

\bTheorem
Consider $G(n,\lambda)$ an $n$-nodes power law graph with $2 < \lambda < 3$. 
The worst case stretch over all node pairs in $G(n,\lambda)$ of a route found by PIE is $O(\log n)$ a.s.
\eTheorem

% This is a worst-case bound when only one tree is used. In the multi-tree extension, smaller trees typically provide good local paths, while larger trees provide good paths across the network.
Note that this is a worst-case bound when only one tree is used.
We observe in the next section that both the average and the maximum stretch do not vary with $n$.

% \subsubsection{Resilience to failures}

\subsection{Protocol Overhead}
% Let us make a few key observations:
We provide a few key observations related to the protocol overhead:
\begin{itemize}
 \item Maintaining a tree requires that each node maintains a shortest path to that tree's root. If $\log n$ trees are used, $\log n$ such shortest paths need to be maintained.
 \item As a comparison, shortest paths algorithms typically rely on distributed protocols that are extremely similar to the spanning tree construction, building $n$ spanning trees, one rooted at each node.
 \item All the control messages used by PIE have a size polylogarithmic in $n$.
 \item More than network overhead, the re-computation of the routing table is perhaps the biggest burden of traditional algorithms. PIE only manipulates extremely small (polylogarithmic) routing state and removes this issue.
 \item For any pro-active routing protocol used in a dynamic topology, there exists a necessary tradeoff between the frequency with which control messages are sent, and the ability of the algorithm to successfully bring packets at destination at any time. We observe in the next section that, due to its geometrical nature, PIE is significantly more resilient to network failures than standard algorithms, and requires to re-compute its state less often.
\end{itemize}

\section{Evaluation}
\label{sec:perfeval}

\subsection{Settings}
We evaluate the behavior of PIE by using simulations on several topologies. 
% To generate larger topologies than allowed by BRITE~\cite{}, 
We wrote our own simulator that we optimized to simulate routing on large graphs. We performed extensive simulations of PIE on several Internet-like graphs~\cite{shavitt:dimes, bu02, barabasi:scaling, bianconi:fit, inet}, on which the results were very similar. To spare space and to be able to explore more of the parameter space, we display results only for the two following topologies:
%  We use graphs obtained both from empirical measures of the Internet topology and using models for the construction of power-law graphs that intend to reproduce the most relevant properties of the Internet. This last category of graphs is useful to assess the scalability aspects of PIE, while the former is useful to assess the behavior of PIE on the best current representations of the Internet. Here are the topologies that we use:
\begin{itemize}
 \item DIMES~\cite{shavitt:dimes} is a collaborative project that uses thousands of end-host agents to reproduce the topology of the Internet as accurately as possible. We use the AS-level dataset of March 2010. We consider all the links as symmetrical and remove the nodes that are not part of the main component, yielding a topology graph of 26,424 AS nodes. We measured $\lambda$ to be about $2.06$ for this graph.
%  \item FIT~\cite{bianconi:fit} is a refinement of the original preferential attachment model proposed by Barab\'{a}si et al.~\cite{barabasi:scaling} that takes into account a ``fitness`` parameter attributed to each node in order to model an evolution process in which all nodes are not equal. While traditional preferential attachment model has been shown to produce power law graphs with exponent $\lambda = 3$, this model produces graphs with $\lambda \sim 2.25$, which is much closer to the Internet node degree distribution.

%  \item GLP (Generalized Linear Preference)~\cite{bu02} is a preferential attachment model that builds on the well-known scheme of Barab\'{a}si et al.~\cite{barabasi:scaling}.
% This model allows us to tune $\lambda$ while producing graphs that exhibit some properties (such as characteristic path length, clustering coefficient or distribution of the highest degrees) close to what is observed in the Internet. To our knowledge, this is one of the most accurate models for generating synthetic Internet-like graphs. The main benefit, of such a synthetic model over a fixed snapshot of the current Internet, is that it allows to generate larger graphs of varying size in order to study the scalability of PIE.
 \item GLP (Generalized Linear Preference)~\cite{bu02} is a preferential attachment model that builds on the well-known scheme of Barab\'{a}si et al.~\cite{barabasi:scaling}.
This model allows us to tune $\lambda$ while producing graphs that 
exhibit some given properties such as characteristic path length, clustering 
coefficient or distribution of the highest degrees. The main benefit, of such a synthetic model over a fixed snapshot of the current Internet, is that it allows to generate larger graphs of varying size in order to study the scalability of PIE.

% We simulate routing on graphs of up to $10^5$ nodes using this model.
% We are able to simulate routing on graphs up to $10^5$ nodes using this model. (\emph{note, I could simulate bigger graphs but that takes more time}).
\end{itemize}
% The FIT and the GLP graphs are generated using a procedure that is exactly similar to the one used by the Brite topology generator~\cite{Brite}.
%We also ran PIE using other graph models~\cite{barabasi:scaling, bianconi:fit, inet}, the results were similar and we do not display them here.

We consider weighted and unweighted graphs. The weights are drawn uniformly in [1,10], which can for instance be thought of as a function of a financial cost and a link capacity, and are comparable to the ISP's link weights range. Such a cost function naturally produces a large amount of violations of the triangle inequality in the graph, which are known to induce much distortion in Euclidean embeddings~\cite{lee:euclidean}.
% As appears in the results, PIE is not handicaped by such a cost function.
For each setting, the statistics have been obtained by simulating routes between $10^5$ distinct, randomly chosen source-destination pairs, over 10 independent experiments using different seeds.
% which amounts to $10^5$ simulated routes per curve point.
% , and typically $10^6$ for a 10-points curve.

\subsection{Results}

% \emph{note: this section can be expended with simulations on larger graphs - 200,000 to $\sim$ 500,000 nodes - using the cluster}

\subsubsection{Performance}
\begin{figure*}
\centering
\begin{tabular}{c c}
PIE & PIE aided by source-routing\\[0.1em]
\includegraphics[width=.475\linewidth]{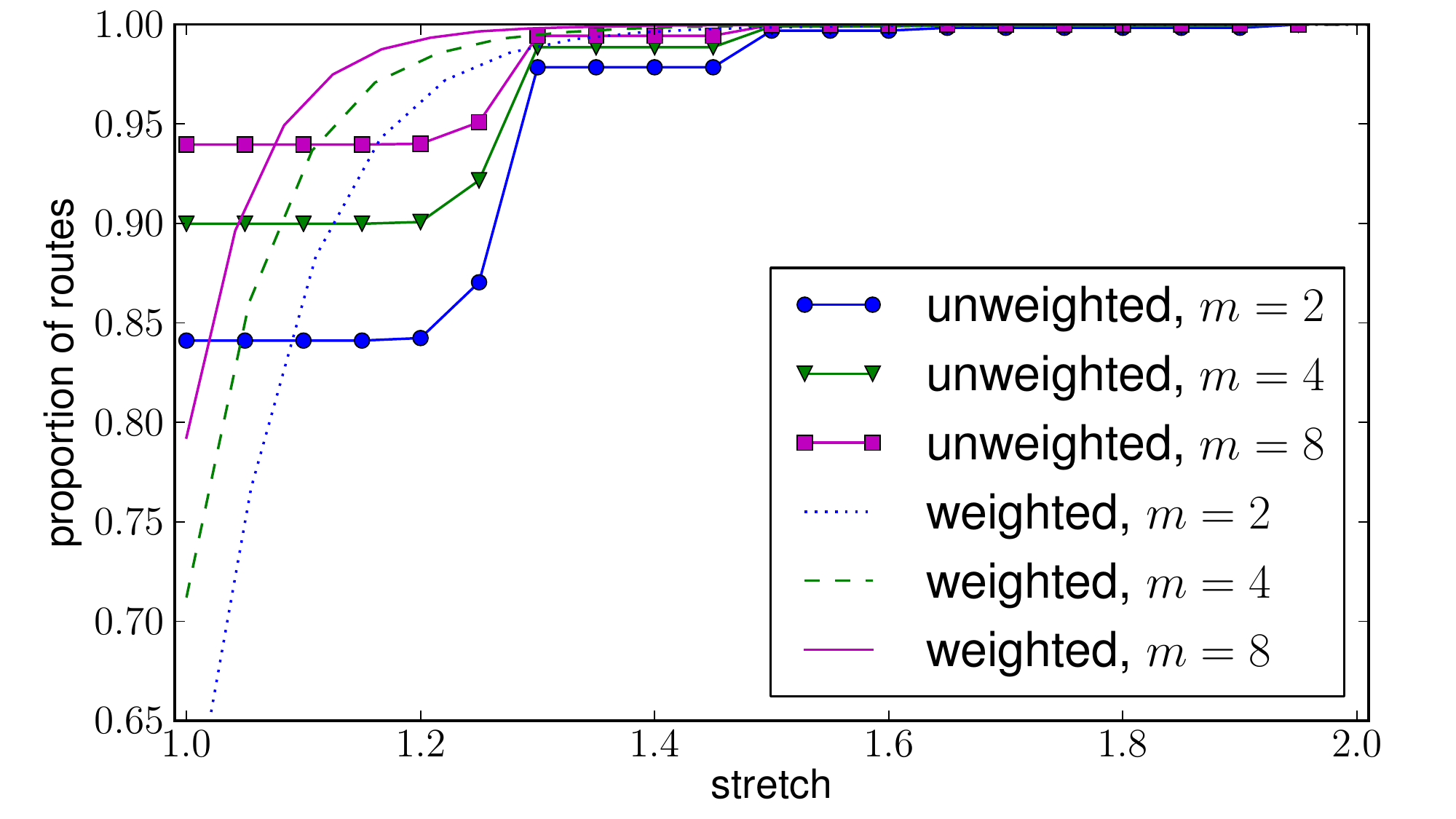} &
\includegraphics[width=.475\linewidth]{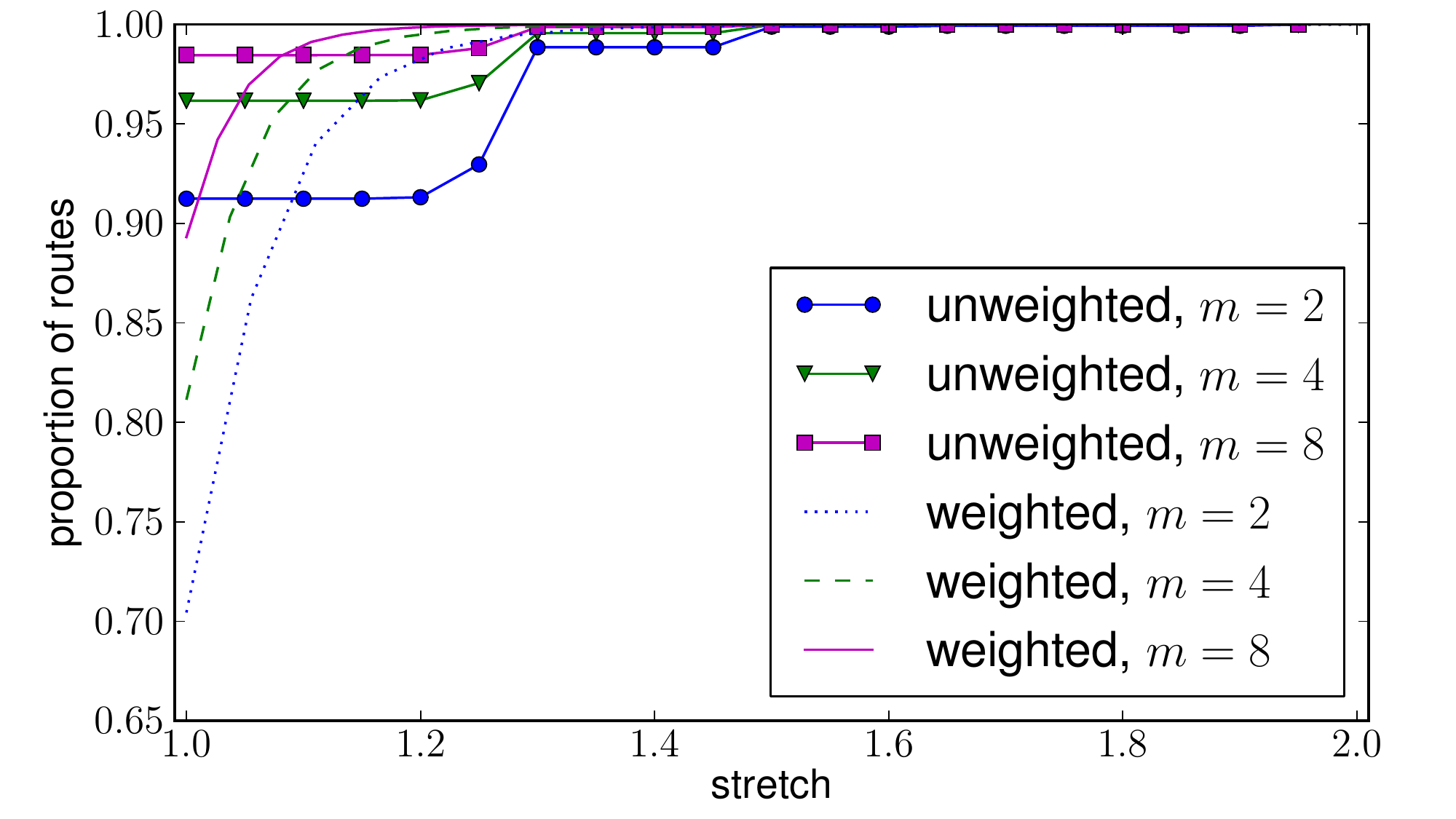}\\
\end{tabular}
\vspace{-1em}
\caption{DIMES topology. Empirical CDFs of the path stretchs for several values of $m$ (the number of levels), with and without costs attributed to links. Left: results for PIE alone. Right: results for PIE aided by the optional source-routing mechanism.}
\vspace{-1.2em}
\label{fig:perfDimes1}
\end{figure*}

\begin{figure}
\centering
\vspace{0.03in}
\includegraphics[width=.94\linewidth]{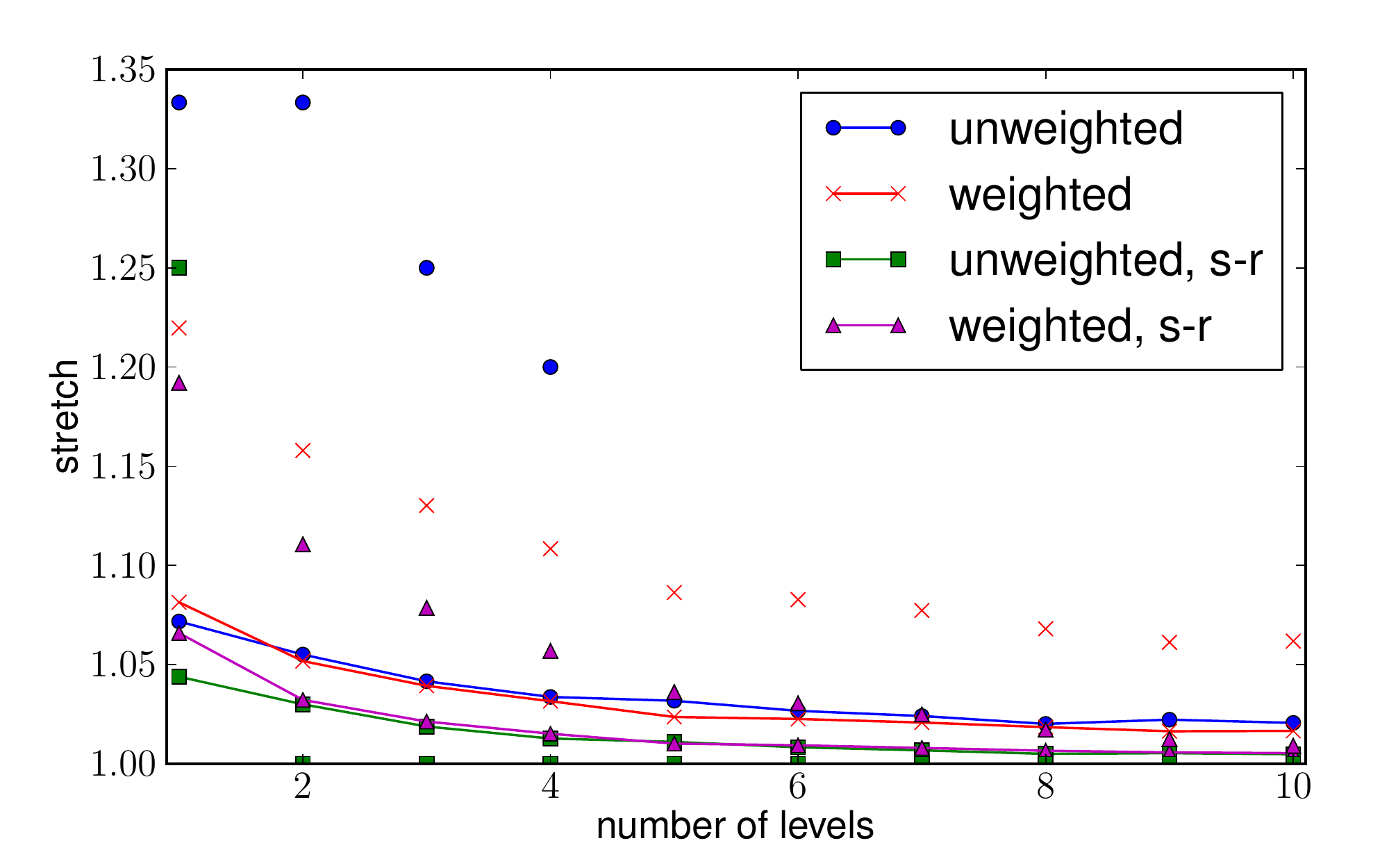}%
\vspace{-0.7em}
\caption{DIMES topology. Stretch as a function of the number of levels $m$, with and without the optional source-routing ("s-r") mechanism. The mean and $90$-th percentile are shown.}
\vspace{-1.2em}
% Middle: maximum stretch as a function of the number of levels.
% Bottom: proportion of shortest routes found by PIE as a function of the number of levels.}
\label{fig:perfDimes2}
\end{figure}

Figure~\ref{fig:perfDimes1} shows the CDFs of the path stretchs obtained on the DIMES topology, and Figure~\ref{fig:perfDimes2} shows the average stretch as a function of the number of locality levels. On both figures, the results are shown both with and without the source-routing extension introduced in Section~\ref{sec:asym}. Even when this extension is not used, the results are excellent: when only two trees are used, more than $97.5$\% of the routes have stretch below $1.3$. For $m \geq 4$, the average stretch is below $1.035$ and, on the unweighted graph, $90$\% or more of the routes found by PIE are the shortest. For $m \geq 8$, the average stretch is below $1.023$. 
%The maximum observed stretch over all the simulations on the weighted graphs was $2.77$ (obtained with $m=1$), and $2$ on the unweighted graph. These values are indeed better than the best possible upper bound of $3$ for compact routing schemes. 
The maximum stretch observed over all the simulations on the unweighted graph was $2$, which is indeed better than the best possible upper bound of $3$ for compact routing schemes.
The source-routing extension further reduces the stretch: with this mechanism enabled, the average stretch is below $1.007$ for both the weighted and unweighted graphs with $m \geq 8$. Unless otherwise specified, we present the remaining results \emph{without} the source-routing extension.
%In the remainder of the paper, the stretch results are shown \emph{without} the source-routing extension.
% any scheme requiring sublinear state.

%The results are striking: the average stretch is below $1.1$ even when only one spanning tree is used, and is even below $1.05$ when each node is part of more than four trees.

%For unweighted graphs, the average stretch goes below 1.05 for 3 levels and more, and below 1.025 for 7 levels and more, while $95$\% of the routes have stretch less than or equal to $4/3$ and the maximum observed stretch (over $10^6$ routes) is $2$, which is below the best possible upper bound of 3 for schemes requiring $o(n)$ state. Moreover, more than $80$\% of the routes found by PIE were the shortest. For weighted graphs, the maximum observed stretch was below 5, and the scheme found between 40\% and 70\% of shortest paths.

\begin{figure}
\centering
\vspace{0.03in}
\includegraphics[width=.94\linewidth]{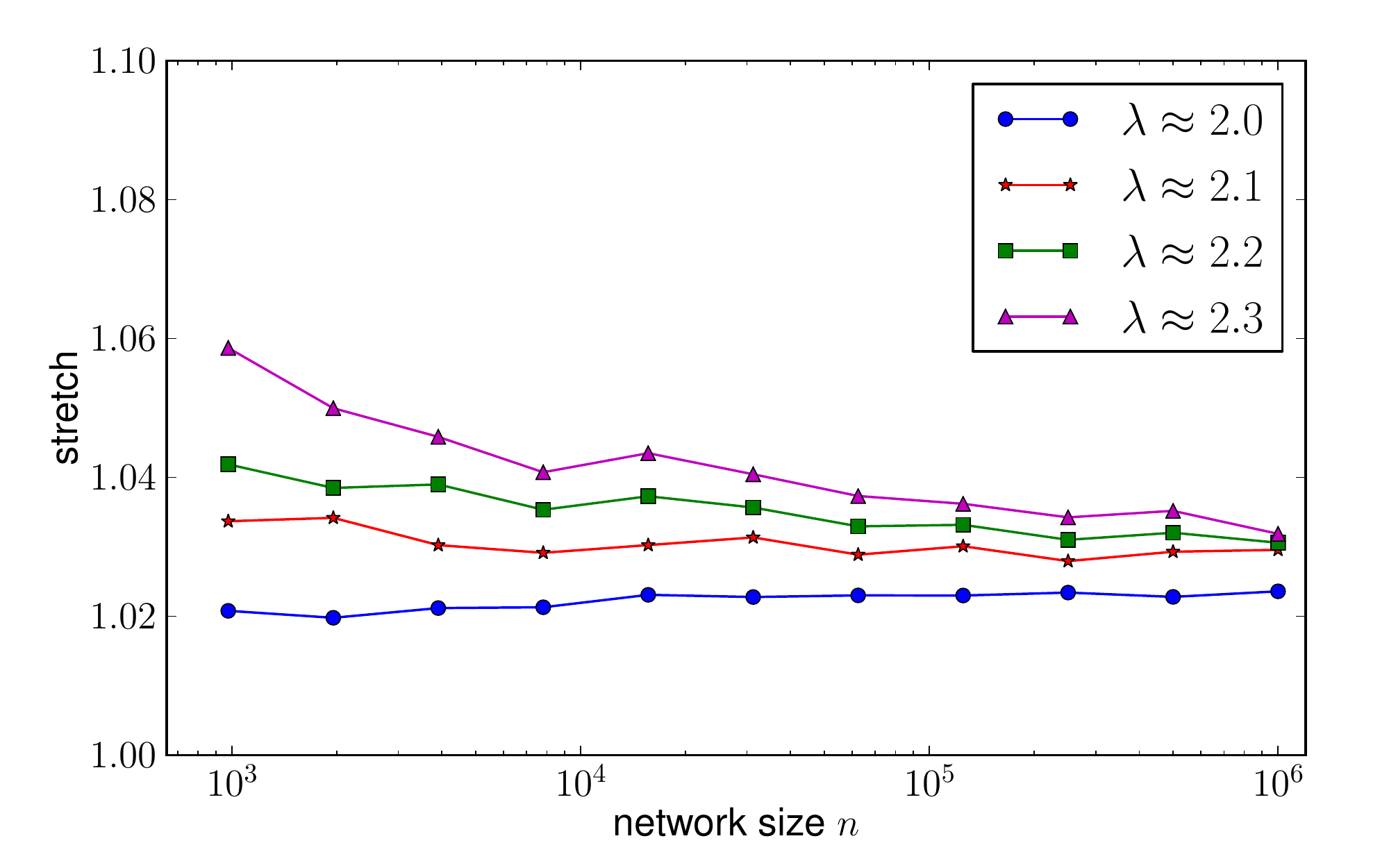}%
% \hfil
% \includegraphics[width=.99\linewidth]{figures/maxStretch02.pdf}%
% \hfil
% \includegraphics[width=.99\linewidth]{figures/propShortest02.pdf}%
\vspace{-0.7em}
\caption{GLP topology. No source routing. Average stretch as a function of $n$, for $2 \leq \lambda \leq 2.3$. For each value of $n$, the $95$-th percentile of the stretch was $4/3$ or less.}
\vspace{-1.2em}
% Middle: maximum stretch as a function of the number of levels.
% Bottom: proportion of shortest routes found by PIE as a function of the number of levels.}
\label{fig:perfGLP1}
\end{figure}

\begin{figure}
\centering
\includegraphics[width=.94\linewidth]{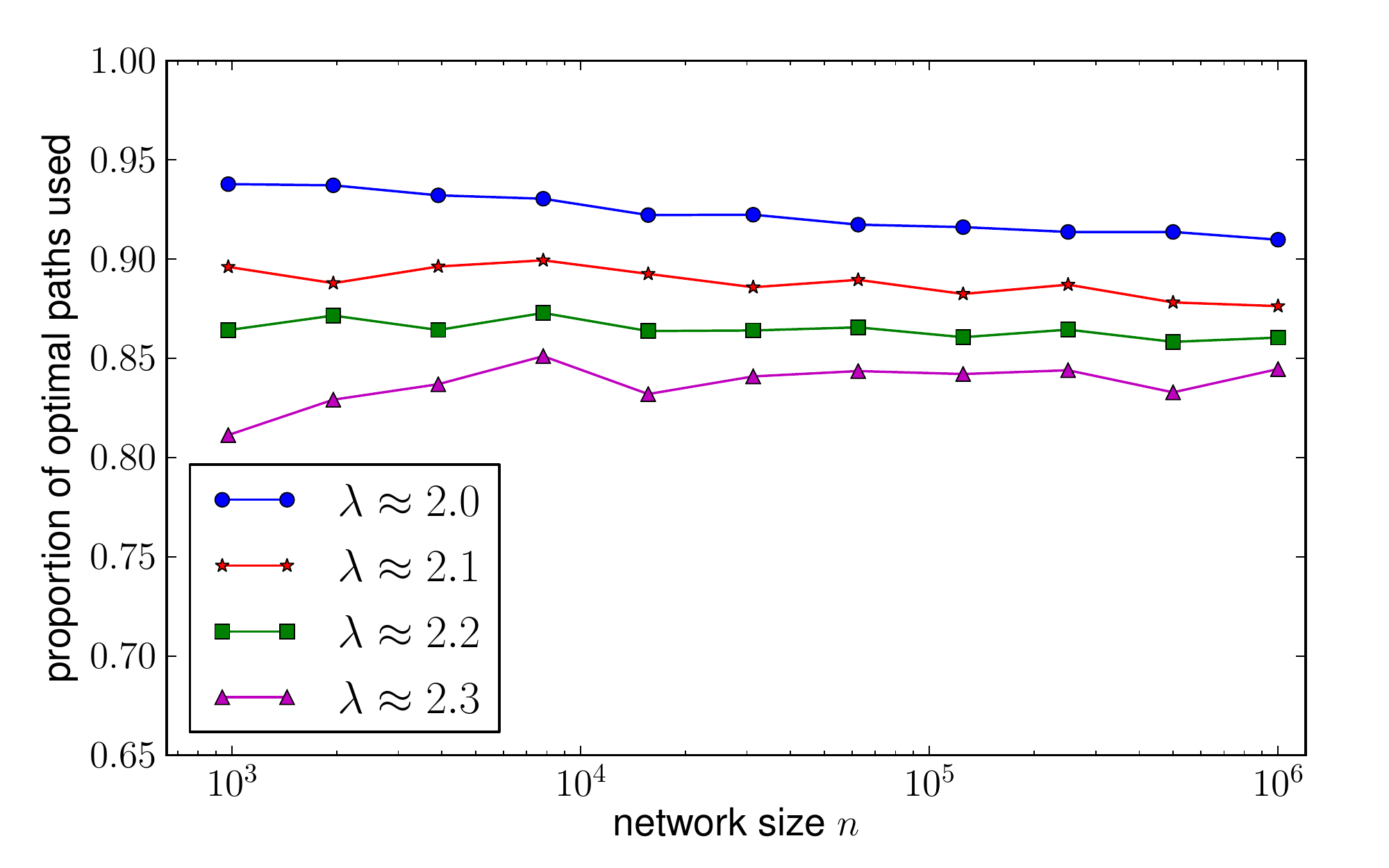}%
\vspace{-0.7em}
\caption{GLP topology. No source routing. Proportion of shortest paths among all the paths found by PIE, as a function of $n$, for $2 \leq \lambda \leq 2.3$.}
\vspace{-1.2em}
\label{fig:perfGLP2}
\end{figure}

Figure~\ref{fig:perfGLP1} shows the evolution of the average stretch when the network grows, on unweighted GLP graphs with $2 \leq \lambda \leq 2.3$. Here and in the following experiments, the number of levels is $m \in O(\log_2 n)$\footnote{The exact function that we use is $m = \lfloor \log_2(n) - 7\rfloor$ (this function yields $m \in \{2,\ldots,12\}$ for the values of $n$ that we consider).}. Figure~\ref{fig:perfGLP2} shows the proportion of shortest paths found by PIE. It appears clearly that the good quality of the routes found by PIE scales perfectly with the size of the network: the average stretch always stays below 1.06 and the proportion of shortest routes above $80$\%. The stretch even appears to slightly decrease with $n$, this comes from our choice for the computation of $m$.

%PIE exhibits similar performances than when evaluated on DIMES, with extremely low stretches, whose average is always below 1.05.

%For similar values of $\lambda$ (although using a different model for the graph generation), the best known compact routing scheme for power-law graphs~\cite{brady} - which does \emph{not} lend itself to a distributed implementation - obtains slightly higher mean stretch (between $\sim1.1$ and $\sim1.21$ for the version achieving polylogarithmic scalability).

\subsubsection{Scalability}
\begin{figure}[t]
% \vspace{-1.5em}
\centering
\includegraphics[width=.94\linewidth]{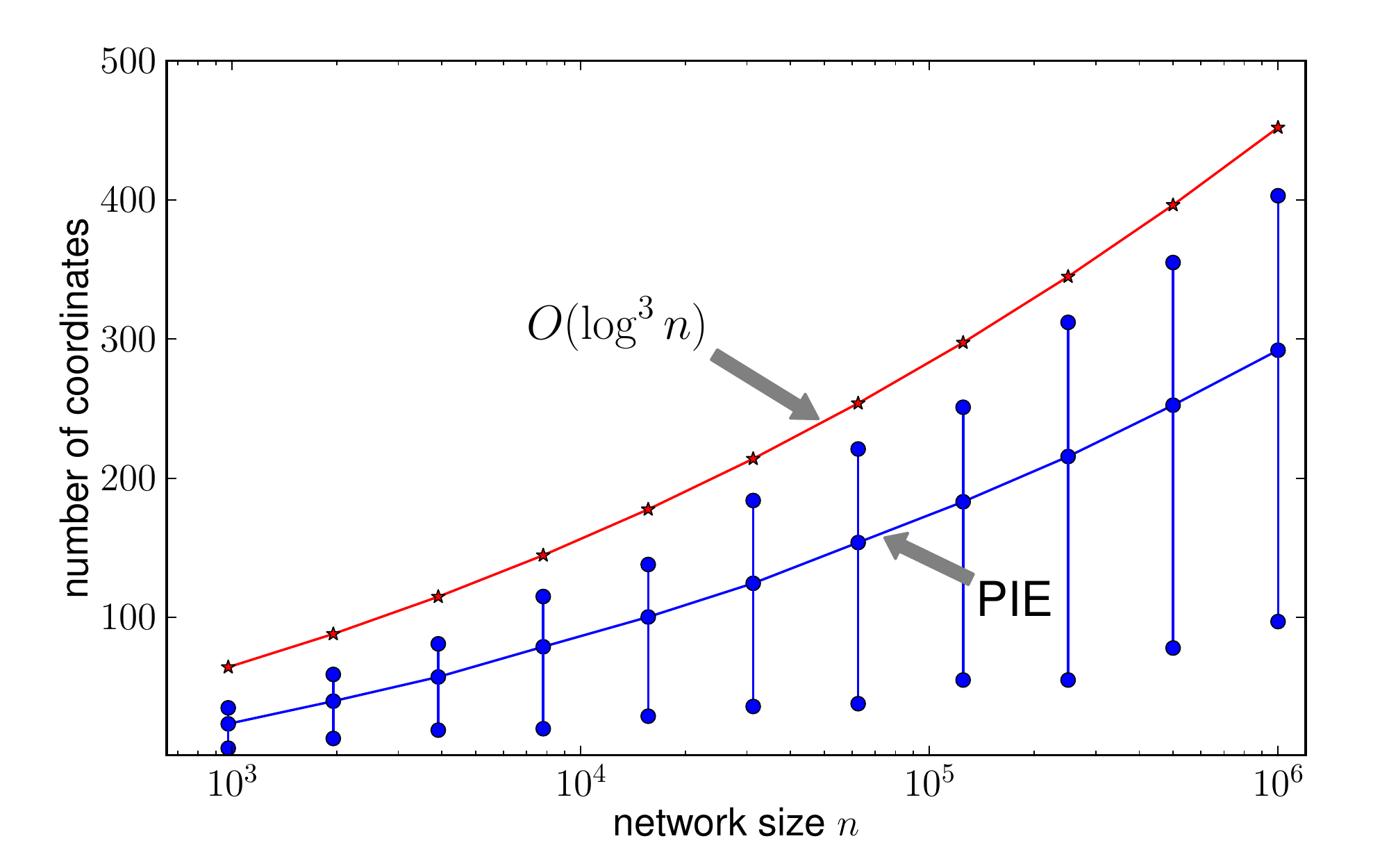}%
\vspace{-1em}
\caption{GLP topology. Scalability of the total number of coordinates. The mininimum, maximum and average number of coordinates per node are shown. The curve above is a (shifted) plot of $(2+\log_{10} n)^3 \in O(\log^3 n)$.}
%\vspace{-1.1em}
\label{fig:scalGLP}
\end{figure}

Figure~\ref{fig:scalGLP} shows the total number of coordinates required at each node by all the trees in the last scenario.
% (for $\lambda \approx 2.1$). 
Also plotted is a (shifted) fit with a function $O(\log^3 n)$ (note the logarithmic scale of the $x$-axis). As predicted in Section~\ref{sec:analysis:scal}, the embedding of $m$ trees by PIE produces $O(\log^3 n)$ coordinates, hence meeting the scalability promises.

\begin{figure}[t]
% \vspace{-1.5em}
\centering
\includegraphics[width=.94\linewidth]{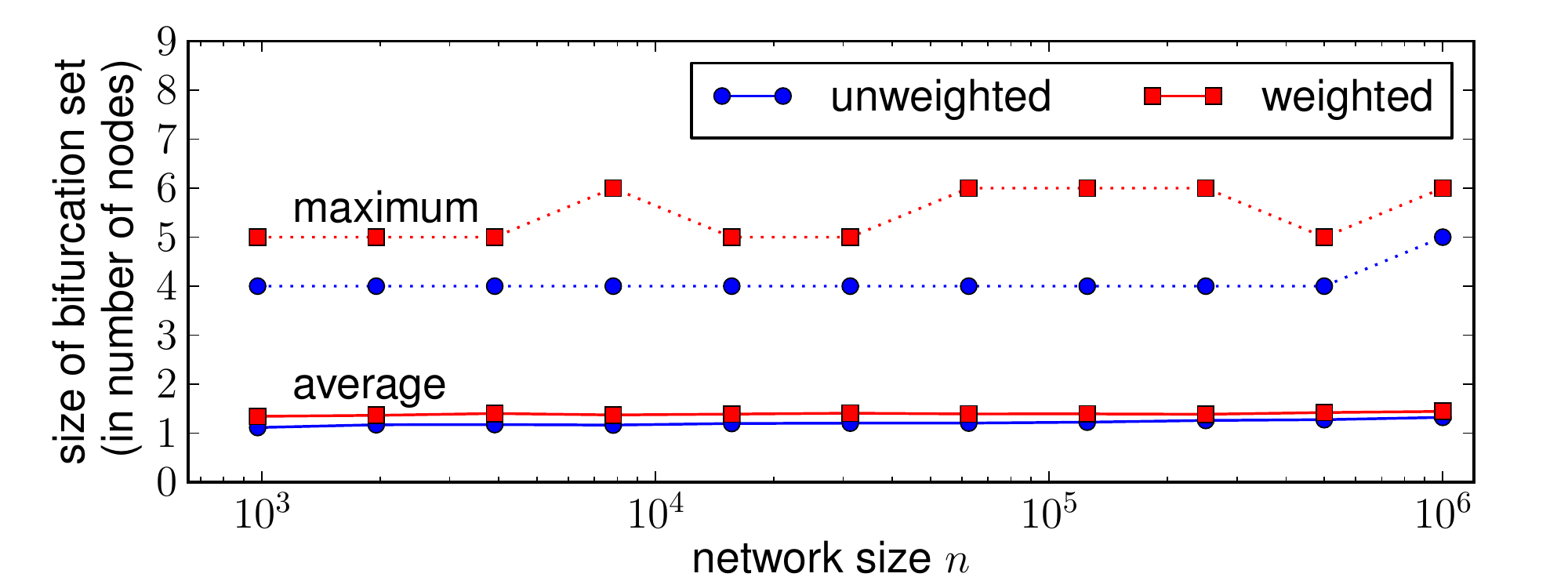}%
\vspace{-1em}
\caption{GLP topology. Size of the bifurcation set used for the optional source-routing mechanism. The average and maximum over $10^5$ routes are shown.}
\vspace{-1.1em}
\label{fig:bif}
\end{figure}

For the optional source-routing extension, Figure~\ref{fig:bif} shows the average and maximum number of bifurcations that need to be included in the packets' headers. We compute the average only over the routes that do benefit from using such a bifurcation set (it would be lower otherwise). 
%Computing the average over all routes would yield much lower values, as a large majority of routes have a bifurcation set size of zero.
Both the average and maximum sizes remain very low -- always $6$ node IDs or less -- and, importantly, they do not grow with the network size. This essentially means that the source-routing extension only incurs a constant overhead in the packets' headers, and the benefits provided in term of routing stretch do not incur a scalability penalty.

%The curve exhibits a growth slightly larger than linear (the $x$-axis has a logarithmic scale), which supports the claim of polylogarithmic scalability (\emph{compare/fit with }$\log^3n$.)

\subsubsection{Resilience to Network Failures}

We consider a scenario in which some randomly chosen nodes fail. If a node has failed, its neighbors cannot send messages to it anymore. We evaluate the success ratio of the routing procedure, after some proportion of the nodes has failed, but before the algorithm has had time to react and adapt the routing tables.

\begin{figure}
% \vspace{-1.5em}
\centering
\includegraphics[width=.94\linewidth]{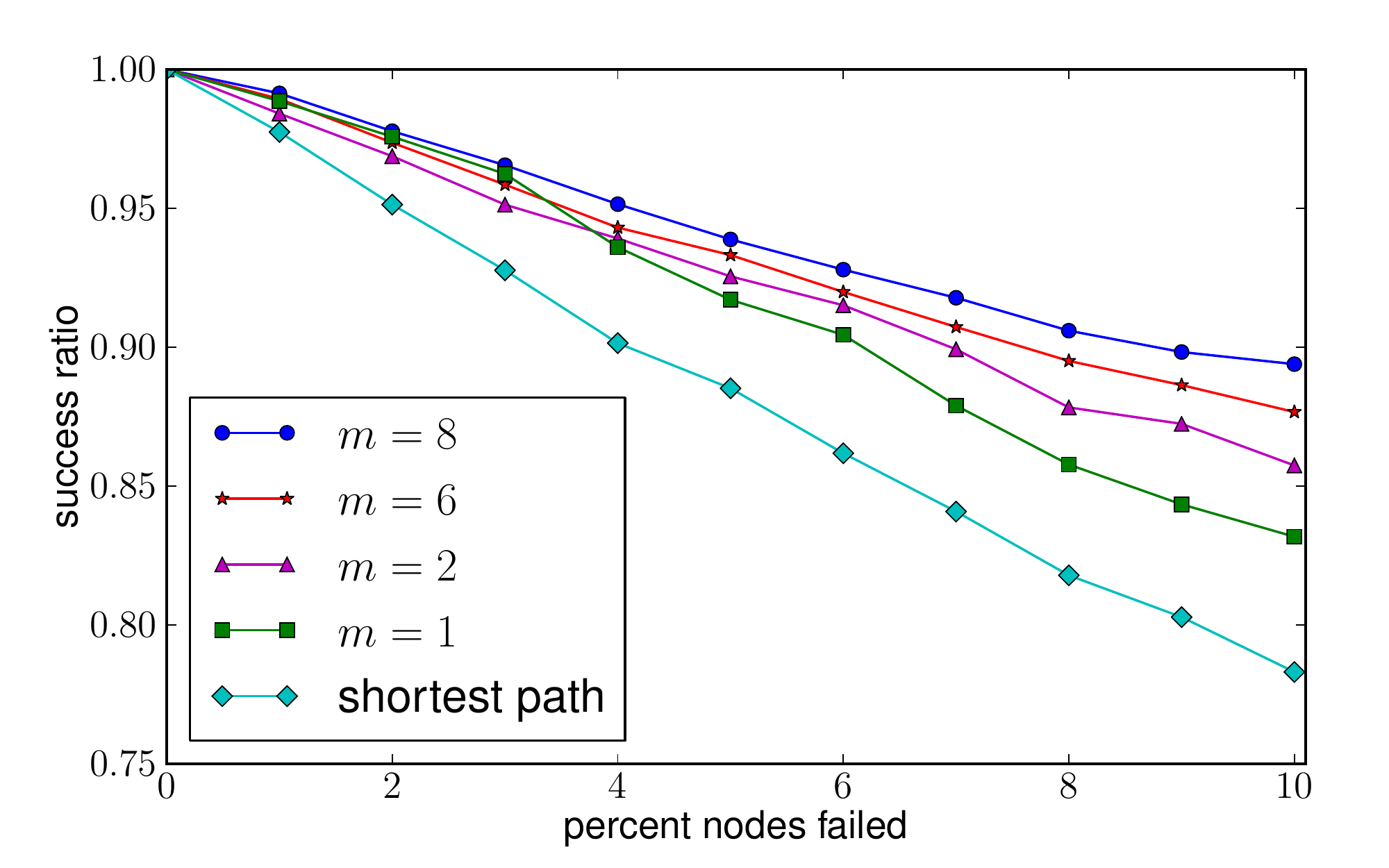}%
\vspace{-1em}
\caption{DIMES topology. Proportion of successful routes as a function of the percentage of failed nodes, for several values of $m$.}
\vspace{-1.3em}
\label{fig:failuresDimes}
\end{figure}

Figure~\ref{fig:failuresDimes} shows the proportion of successful routes (success ratio) as a function of the percentage of nodes that have failed, on the unweighted DIMES topology. PIE maintains a significantly higher number of successful paths than traditional shortest paths algorithms. This is explained by considering the greedy nature of the forwarding procedure of PIE: forwarding to \emph{any} neighbor that is closer to the destination provides
route diversity gain, while schemes producing $1$-to-$1$ mappings between destinations and next hops (including compact routing schemes), do not benefit from this route diversity.

When only one tree is used, PIE consistently reduces the number of routing failures by at least 20\%, and this proportion jumps to 50\% when $m = 8$.
Even in the extremely unlikely scenario where 10\% of the Internet fails, PIE could manage to maintain a high success ratio, about 90\%.
One direct consequence is that, for a given success ratio, PIE does not need to recompute its state as often as standard algorithms.
% One direct consequence is that PIE does not need to recompute its state as often as standard algorithms 
% %and thus produces less network overhead 
% to maintain similar success ratio.

\subsubsection{Comparison with the State of the Art}
We compare PIE with the \emph{general} compact routing scheme~\cite{thorup:compact} (that we denote by TZ). In addition we also compare it with the \emph{specialized} compact routing scheme~\cite{brady}, that is especially targeted for power law graphs (that we denote by BC). It is proposed in~\cite{brady} to combine TZ and BC in order to obtain a new scheme (that we denote by TZ+BC), which uses the best of the routes found by TZ and BC taken together. We recall that TZ achieves only $O(\sqrt{n\log n})$ scalability for the routing table size, and BC requires a complete knowledge of the graph at all the nodes and is not translated in a distributed protocol. TZ+BC accumulates these two fundamental issues.

\begin{figure}
% \vspace{-1.3em}
\centering
\includegraphics[width=.94\linewidth]{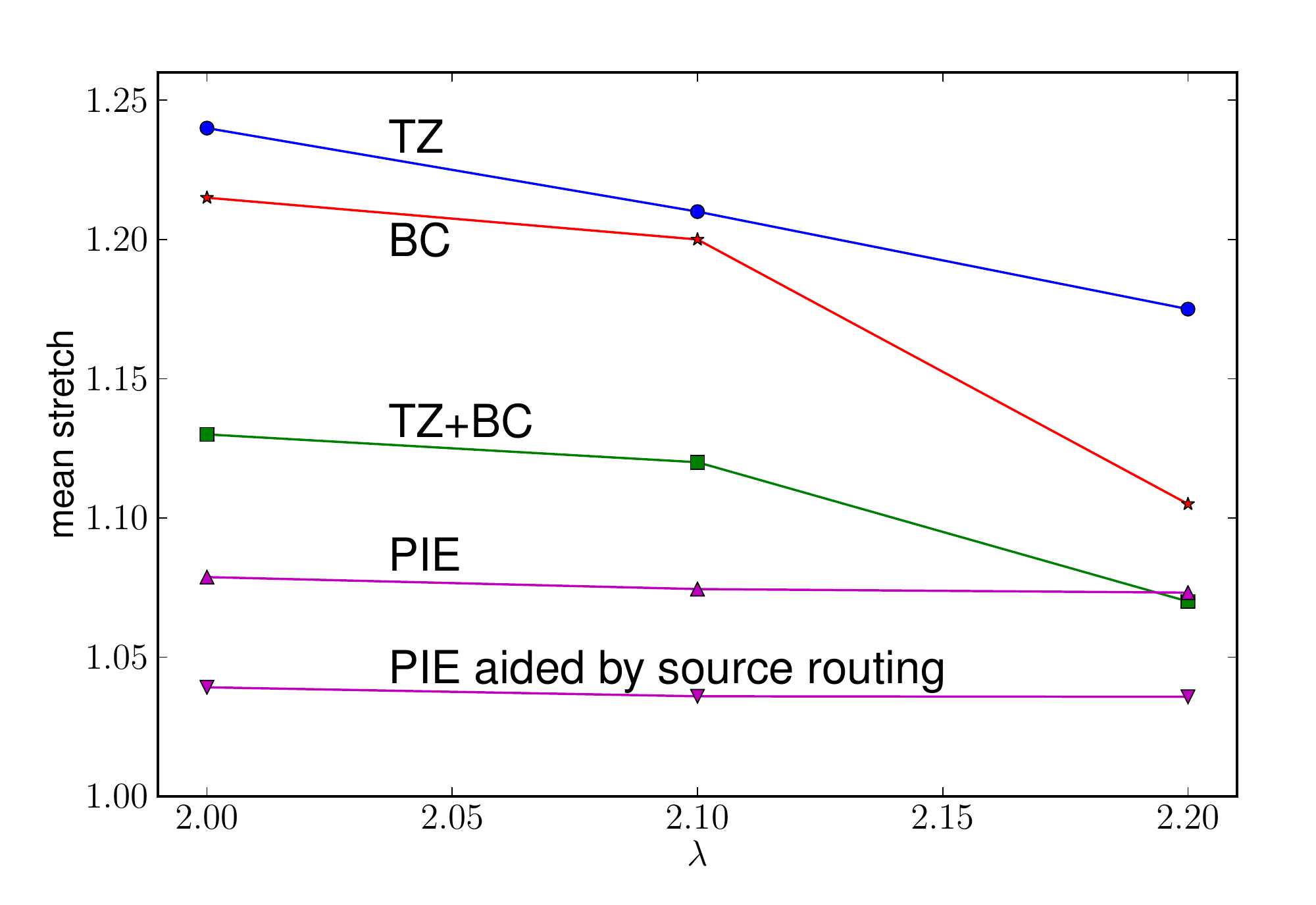}%
\vspace{-1em}
\caption{Comparison of the mean stretch obtained by PIE,~\cite{thorup:compact} (TZ) and~\cite{brady} (BC).
%  as a function of $\lambda$. 
The values plotted for BC and TZ come from~\cite{brady}, as do the graphs used for the simulations. In this scenario, the number of nodes is $10^4$, but the main connected components of the graphs have size $n \approx 8400$.}
% ~\cite{brady} provides values only for $\lambda \in \{2.0, 2.1, 2.2\}$.}
\vspace{-1.2em}
% Note that no values are reported in~\cite{brady} for $\lambda > 2.2$.}
\label{fig:comp}
\end{figure}

The authors of~\cite{brady} publicly provide the graphs that they used to obtain their simulation results with $\lambda  \in \{2,2.1,2.2\}$.
We can therefore run PIE (with and without the optional source-routing extension proposed in Section~\ref{sec:asym}) on these exact same graphs and compare the results. This is shown on Figure~\ref{fig:comp}. Critically, PIE performs significantly better than its less scalable, respectively centralized, counterparts. It even finds similar or better routes than the best routes found by TZ and BC taken together.

\section{Discussion}
\label{sec:discussion}
% While we demonstrate the scalability of our routing method from a
% theoretical point of view and its practicality for implementation,
% translating our protocol to a deployment environment requires a couple
% more steps.
While we demonstrate the scalability of our routing method from a
theoretical point of view and provide the corresponding distributed protocol,
translating this protocol to a deployment environment requires a couple
more steps.
In a single administrative domain, its deployment would be
easy, as ASs run their own routing protocol internally. Since some ASs
are relatively large, they would benefit from the scalability of our
scheme.
Similarly, our protocol would be practical over large overlay
networks, where the weights would be dependent on the target that the
overlay aims to achieve (for instance, minimize delay between overlay nodes).

For the wider Internet, the issue becomes to integrate our protocol
with BGP.
The simplest integration would be to build tree(s) of level $0$ between the ASs and trees of higher levels within the ASs. Internally, the ease of geometric routing would prevail. Externally, BGP tables and the existing IP nomenclature could be kept. Such an approach would already benefit from the lightweight geometric coordinates for forwarding, but would still require $O(n)$ memory.
% to store explicit per-destination policies.

% The simplest integration would be to have one or several trees of level
% $0$ built using BGP as currently, and have the trees at the other
% locality levels built into each AS. Since trees in between ASs would
% have no common nodes, BGP policies would strictly apply in between
% ASs, while internally, the ease of routing on geographic coordinates
% would prevail. However, this increases the size of the routing state,
% if only a little, as it would still scale as $O(n)$.

The best integration would thus be to modify BGP so as to fully take
advantage of PIE. While this is beyond the scope of this paper, we
contend that it is possible to achieve. As a simple example, consider four
ASs, AS1 through AS4, with AS1 and AS4 both being connected through
both AS2 and AS3. Assume that BGP is configured so as to prevent AS3
from being used as a transit AS between AS1 and AS4. Assume further
that, rather than using STP to create and propagate the coordinates,
we now use a BGP-like mechanism.

When AS3 receives the eBGP message from AS1 to create routing
coordinates, it propagates it internally, but not through its eBGP
connection to AS4. On the other hand, AS2 does according to its BGP
policy. Thus, traffic from AS1 to AS4 will see AS2 as in between them in the metric space,
and AS3 as in a wrong direction and routing will naturally go through
AS2. The weights between ASs can be built upon the BGP attributes
as well.
The fact that PIE adapts well to arbitrary link costs
provides good support to use it for traffic engineering.

This basic example shows that there is enough expressiveness in
creating the coordinates of PIE to satisfy some basic policy
mechanisms.

\section{Conclusion}
\label{sec:conclusion}
% Remains to do: policies

We have presented and evaluated PIE, a distributed protocol that produces a greedy embedding. It does so by isometrically embedding trees in non-Euclidean spaces of dimension $O(\log^2(n))$. Each node in the graph belongs to $O(\log(n))$ trees.
%and the trees are built in a way that fits well the self-similar tree structure of Internet-like graphs. 
The greediness of the embedding allows the forwarding procedure to take any available shortcut off the trees while avoiding loops and guaranteeing the success of routing.
PIE typically relaxes the \emph{deterministic} guarantees provided by classic compact routing schemes, in order to be written as a distributed protocol. The bottomline of the good features of PIE is that these guarantees are now \emph{probabilistic}, satisfied \emph{with probability one} on the relevant categories of large graphs.
%The theory behind the good features of PIE relies on the very strong statistical properties of Internet-like graphs: It is possible to build a graph on which PIE requires $O(n)$ state, but for large Internet-like graphs, as well as for most random graphs, the probability to meet such a graph is zero.

We have proved that PIE achieves a success ratio of 100\% on any graph, that it provides polylogarithmic scalability, and we have given a logarithmic upper bound on the path stretch. We have used large-scale simulation on synthetic and real-world topologies to observe that the stretch is independent of $n$ and that it remains extremely low, typically lower than for centralized or less scalable state-of-the-art algorithms. In addition, we have proposed an optional source-aided routing mechanism that provides significant stretch improvement with no scalability penalty.

%We also tested PIE on unit disc graphs, with good results compared to traditional geometric routing algorithms, suggesting avenues for ad-hoc/wireless networks as well.

PIE comes in a clean-slate perspective. We briefly discussed the challenges related to any replacement of the existing protocols and gave indications that such a geometric scheme could be used with traffic engineering and policy routing, making this a direction worth exploring for future work.
%  An interesting direction for future work would thus consist in exploring this direction further.
% We however gave some indications that the adaptation of the geometric scheme for this purpose is a direction worth exploring.
% PIE comes in a clean-slate perspective, and we do not consider the numerous subtleties inherent to a protocol like BGP. It would be an interesting direction for future work to study how several scalable coordinates systems that fit the AS structure of the Internet could be used to implement features such as policy routing and traffic engineering. 

% The fact that PIE adapts very well to arbitrary link costs gives good support in pursuance of this direction.
% As an orthogonal consideration, the good stretch performance obtained by PIE is
We can draw a few orthogonal considerations from the good stretch performance obtained by PIE. It is 
a direct indicator of the self-similar tree-like structure of the Internet,
%The very low path stretchs obtained even on weighted graphs seem to be strong indicators that 
and it shows that the embedding has low distortion. It would thus probably suit well distance estimation tasks in the Internet, 
as it is required by many overlay and peer-to-peer applications.
%We however did not conduct any evaluation in this direction.

\bibliographystyle{IEEEtran}
\bibliography{geobib}

\begin{thebibliography}{10}
\providecommand{\url}[1]{#1}
\csname url@samestyle\endcsname
\providecommand{\newblock}{\relax}
\providecommand{\bibinfo}[2]{#2}
\providecommand{\BIBentrySTDinterwordspacing}{\spaceskip=0pt\relax}
\providecommand{\BIBentryALTinterwordstretchfactor}{4}
\providecommand{\BIBentryALTinterwordspacing}{\spaceskip=\fontdimen2\font plus
\BIBentryALTinterwordstretchfactor\fontdimen3\font minus
  \fontdimen4\font\relax}
\providecommand{\BIBforeignlanguage}[2]{{%
\expandafter\ifx\csname l@#1\endcsname\relax
\typeout{** WARNING: IEEEtran.bst: No hyphenation pattern has been}%
\typeout{** loaded for the language `#1'. Using the pattern for}%
\typeout{** the default language instead.}%
\else
\language=\csname l@#1\endcsname
\fi
#2}}
\providecommand{\BIBdecl}{\relax}
\BIBdecl

\bibitem{herzen:pie}
J.~Herzen, C.~Westphal, and P.~Thiran, ``Scalable routing easy as pie: A
  practical isometric embedding protocol,'' in \emph{Network Protocols (ICNP),
  2011 19th IEEE International Conference on}, 2011, pp. 49--58.

\bibitem{iab}
D.~Meyer, L.~Zhang, and K.~Fall, ``Report from the iab workshop on routing and
  addressing,'' \url{http://tools.ietf.org/html/draft-iab-raws-report-01.html},
  Feb. 2007.

\bibitem{gavoille97}
C.~Gavoille and M.~Gengler, ``Space-efficiency for routing schemes of stretch
  factor three,'' \emph{Journal of Parallel and Distributed Computing},
  vol.~61, pp. 61--679, 1997.

\bibitem{milgram67smallworld}
S.~Milgram, ``The small world problem,'' \emph{Psychology Today}, vol.~2, pp.
  60--67, 1967.

\bibitem{pastor-satorras04}
R.~Pastor-Satorras and A.~Vespignani, \emph{Evolution and Structure of the
  Internet: A Statistical Physics Approach}.\hskip 1em plus 0.5em minus
  0.4em\relax New York, NY, USA: Cambridge University Press, 2004.

\bibitem{papadimitriou:conjecture}
C.~Papadimitriou and D.~Ratajczak, ``On a conjecture related to geometric
  routing,'' \emph{Theoretical Computer Science}, vol. 244, no.~1, 2005.

\bibitem{rao:geographic}
A.~Rao, S.~Ratnasamy, C.~Papadimitriou, S.~Shenker, and I.~Stoica, ``Geographic
  routing without location information,'' in \emph{Proceedings of ACM MobiCom},
  2003, pp. 96--108.

\bibitem{Dabek:vivaldi}
F.~Dabek, R.~Cox, F.~Kaashoek, and R.~Morris, ``Vivaldi: a decentralized
  network coordinate system,'' \emph{SIGCOMM Comput. Commun. Rev.}, vol.~34,
  no.~4, pp. 15--26, 2004.

\bibitem{goafr}
F.~Kuhn, R.~Wattenhofer, Y.~Zhang, and A.~Zollinger, ``Geometric ad-hoc
  routing: of theory and practice,'' in \emph{PODC '03}, 2003, pp. 63--72.

\bibitem{durocher:3d}
S.~Durocher, D.~Kirkpatrick, and L.~Naranayan, ``On routing with guaranteed
  delivery in three-dimensional ad hoc wireless newtorks,'' in
  \emph{Proceedings of ICDNC}, 2008, pp. 546--557.

\bibitem{subramanian:optimal}
S.~Subramanian, S.~Shakkottai, and P.~Gupta, ``On optimal geographic routing in
  wireless networks with holes and non-uniform traffic,'' in \emph{Proceedings
  of Infocom}, 2008.

\bibitem{matousek}
P.~Indyk and J.~Matousek, ``Low-distortion embeddings of finite metric
  spaces,'' in \emph{in Handbook of Discrete and Computational Geometry}.\hskip
  1em plus 0.5em minus 0.4em\relax CRC Press, 2004, pp. 177--196.

\bibitem{maymounkov:greedy}
P.~Maymounkov, ``Greedy embeddings, trees and {E}uclidian vs. {L}obachevsky
  geometry,'' Technical Report, available at
  http://pdos.csail.mit.edu/~petar/pubs.html, 2006.

\bibitem{moitra:greedy}
A.~Moitra and T.~Leighton, ``Some results on greedy embeddings in metric
  spaces,'' \emph{Foundations of Computer Science, Annual IEEE Symposium on},
  vol.~0, pp. 337--346, 2008.

\bibitem{kleinberg:hyperbolic}
R.~Kleinberg, ``Geographic routing using hyperbolic space,'' in
  \emph{Proceedings of Infocom}, 2007.

\bibitem{Crovella:hyperbolic}
A.~Cvetkovski and M.~Crovella, ``Hyperbolic embedding and routing for dynamic
  graphs,'' in \emph{Proceedings of Infocom 2009}, April 2009.

\bibitem{papadopoulos10}
M.~B. Fragkiskos~Papadopoulos, Dmitri~Krioukov and A.~Vahdat, ``Greedy
  forwarding in dynamic scale-free networks embedded in hyperbolic metric
  spaces,'' in \emph{Proceedings of Infocom}, 2010.

\bibitem{Krioukov:sustaining}
M.~Bogu\~{n}\'{a}, F.~Papadopoulos, and D.~Krioukov, ``{Sustaining the Internet
  with hyperbolic mapping},'' \emph{Nature Communications}, vol.~1, no.~6, pp.
  1--8, September 2010.

\bibitem{Pei:greedy}
C.~Westphal and G.~Pei, ``Scalable routing via greedy embedding,'' in
  \emph{Proceedings of Infocom Mini-Conference}, April 2009.

\bibitem{gupta}
A.~Gupta, A.~Kumar, and R.~Rastogi, ``Traveling with a pez dispenser (or,
  routing issues in mpls),'' \emph{SIAM J. Comput.}, vol.~34, no.~2, 2005.

\bibitem{greedy:flury}
R.~Flury, S.~Pemmaraju, and R.~Wattenhofer, ``Greedy routing with bounded
  stretch,'' in \emph{Proc. of Infocom}, April 2009.

\bibitem{krioukov:sigcomm}
D.~Krioukov, kc~claffy, K.~Fall, and A.~Brady, ``On compact routing for the
  {I}nternet,'' \emph{SIGCOMM Comp. Comm. Rev.}, vol.~37, no.~3, 2007.

\bibitem{thorup:compact}
M.~Thorup and U.~Zwick, ``Compact routing schemes,'' in \emph{{ACM} Symposium
  on Parallel Algorithms and Architectures}, 2001, pp. 1--10.

\bibitem{mao:S4}
Y.~Mao, F.~Wang, L.~Qiu, S.~Lam, and J.~Smith, ``S4: Small state and small
  stretch routing protocol for large wireless sensor networks,'' in
  \emph{Proceedings of the 4th USENIX NSDI 2007}, April 2007.

\bibitem{singla2010}
A.~Singla, P.~B. Godfrey, K.~Fall, G.~Iannaccone, and S.~Ratnasamy, ``Scalable
  routing on flat names,'' in \emph{Proceedings of Co-NEXT}, 2010.

\bibitem{Chen:compact}
W.~Chen, C.~Sommer, S.-H. Teng, and Y.~Wang, ``Compact routing in power-law
  graphs,'' in \emph{Proceedings of DISC'09}, 2009, pp. 379--391.

\bibitem{brady}
A.~Brady and L.~Cowen, ``Compact routing on power-law graphs with additive
  stretch,'' in \emph{ALENEX}, 2006.

\bibitem{caesar:VRR}
M.~Caesar, M.~Castro, E.~Nightingale, G.~O'Shea, and A.~Rowstron, ``{Virtual
  Ring Routing}: Network routing inspired by {DHT}s,'' in \emph{Proc. of ACM
  SIGCOMM'06}, 2006, pp. 351--362.

\bibitem{ghaffari:delaunay}
M.~Ghaffari, B.~Hariri, and S.~Shirmohammadi, ``On the necessity of using
  {Delaunay} triangulation substrate in greedy routing based networks,''
  \emph{IEEE Communications Letters}, vol.~14, no.~3, pp. 266--268, March 2010.

\bibitem{lee:euclidean}
S.~Lee, Z.-L. Zhang, S.~Sahu, and D.~Saha, ``On suitability of {E}uclidean
  embedding for host-based network coordinate systems,'' \emph{Networking,
  IEEE/ACM Transactions on}, vol.~18, no.~1, pp. 27 --40, 2010.

\bibitem{linial}
N.~Linial, E.~London, and Y.~Rabinovich, ``The geometry of graphs and some of
  its algorithmic applications,'' \emph{Combinatorica}, vol.~15, pp. 577--591,
  1994.

\bibitem{perlman:stp}
R.~Perlman, ``An algorithm for distributed computation of a spanning tree in an
  extended {LAN},'' \emph{ACM SIGCOMM Computer Communication Review}, vol.~15,
  no.~4, pp. 44--53, 1985.

\bibitem{faloutsos99}
M.~Faloutsos, P.~Faloutsos, and C.~Faloutsos, ``On power-law relationships of
  the internet topology,'' in \emph{Proceedings of SIGCOMM '99}.\hskip 1em plus
  0.5em minus 0.4em\relax New York, NY, USA: ACM, 1999, pp. 251--262.

\bibitem{mahadevan06}
P.~Mahadevan, D.~Krioukov, M.~Fomenkov, X.~Dimitropoulos, k.~c. claffy, and
  A.~Vahdat, ``The internet as-level topology: three data sources and one
  definitive metric,'' \emph{SIGCOMM Comp. Comm. Rev.}, 2006.

\bibitem{Chung02theaverage}
F.~Chung and L.~Lu, ``The average distances in random graphs with given
  expected degrees,'' \emph{Internet Mathematics}, vol.~1, pp.
  15\,879--15\,882, 2002.

\bibitem{li:scalefree}
L.~Li, D.~Alderson, J.~C. Doyle, and W.~Willinger, ``Towards a theory of
  scale-free graphs: Definition, properties, and implications,'' \emph{Internet
  Mathematics}, vol.~2, p.~4, 2005.

\bibitem{lu:diameter}
F.~Chung and L.~Lu, ``The diameter of random sparse graphs,'' in \emph{Advances
  in Applied Math}, pp. 257--279.

\bibitem{shavitt:dimes}
Y.~Shavitt and E.~Shir, ``Dimes: let the internet measure itself,''
  \emph{SIGCOMM Comput. Commun. Rev.}, vol.~35, pp. 71--74, October 2005.

\bibitem{bu02}
T.~Bu and D.~F. Towsley, ``On distinguishing between internet power law
  topology generators,'' in \emph{Proc. of Infocom}, 2002.

\bibitem{barabasi:scaling}
A.-L. Barabasi and R.~Albert, ``Emergence of scaling in random networks,''
  \emph{Science}, vol. 286, no. 5439, pp. 509--512, 1999.

\bibitem{bianconi:fit}
G.~Bianconi and A.-L. Barabási, ``Competition and multiscaling in evolving
  networks,'' \emph{Europhysics Letters}, vol.~54, no.~4, p. 436, 2001.

\bibitem{inet}
J.~Winick and S.~Jamin, ``Inet-3.0: Internet topology generator,'' Tech. Rep.,
  2002.

\end{thebibliography}

\end{document}